\documentclass[fleqn,usenatbib]{mnras}

\usepackage{graphicx}
\usepackage{hyperref}

\usepackage{newtxtext,newtxmath}
\usepackage[T1]{fontenc}

\usepackage{amsmath}

\usepackage{xspace}
\usepackage{gensymb}
\usepackage{subfigure}
\usepackage{color}

\DeclareRobustCommand{\VAN}[3]{#2}
\let\VANthebibliography\thebibliography
\def\thebibliography{\DeclareRobustCommand{\VAN}[3]{##3}\VANthebibliography}

\usepackage[encapsulated]{CJK}
\usepackage[utf8x]{inputenc}

\newcommand{\cotwoone}{CO(2-1)\xspace}

\newcommand{\coonezero}{CO(1-0)\xspace}

\newcommand{\kks}{${\rm K\,km\,s^{-1}}$\xspace}

\newcommand{\msol}{M$_{\odot}$\xspace}

\newcommand{\kms}{${\rm km\,s^{-1}}$\xspace}
\newcommand{\htwo}{H$_{2}$\xspace}
\newcommand{\msolpcsq}{M$_{\odot}\,{\rm pc}^{-2}$\xspace}

\newcommand{\msolconvert}{M$_{\odot}\,{\rm pc^{-2}/\,\rm K\,km\,s^{-1}}$\xspace}

\newcommand{\cmtwo}{cm$^{-2}$\xspace}
\newcommand{\hi}{{\sc H\,i}\xspace}

\newcommand{\hibold}{\bfseries{\scshape{H\,i}}\xspace}

\newcommand{\ci}{{\sc C\,i}\xspace}
\newcommand{\cii}{{\sc C\,ii}\xspace}

\newcommand{\tspin}{$T_{\rm s}$\xspace}

\newcommand{\tpeak}{$T_{\rm p}$\xspace}

\title[\hi spectra prefer a multi-component model]{A lack of constraints on the cold opaque \hibold mass: \hibold spectra in M31 and M33 prefer multi-component models over a single cold opaque component}

\author[Koch et al.]{Eric W. Koch$^{1,2}$\thanks{E-mail:
koch.eric.w@gmail.com (EWK)},
Erik W. Rosolowsky$^{1}$,
Adam K. Leroy$^{3}$,
J\'er\'emy Chastenet$^{4}$,
\newauthor
I-Da Chiang \begin{CJK*}{UTF8}{bkai}(江宜達)\end{CJK*}$^{4}$,
Julianne Dalcanton$^{6}$,
Amanda A. Kepley$^{7}$,
Karin M. Sandstrom$^{4}$,
\newauthor
Andreas Schruba$^{8}$,
Sne\v{z}ana Stanimirovi\'{c}$^{9}$,
Dyas Utomo$^{3,7}$,
Thomas G. Williams$^{10}$
\\
$^{1}$University of Alberta, Department of Physics, 4-183 CCIS, Edmonton AB T6G 2E1, Canada\\
$^{2}$Center for Astrophysics $\mid$ Harvard \& Smithsonian, 
60 Garden St., Cambridge, MA 02138, USA\\
$^{3}$The Ohio State University, Department of Astronomy, 140 West 18th Avenue, Columbus, OH 43210, USA\\
$^{4}$Center for Astrophysics and Space Sciences, Department of Physics, University of California, San Diego, 9500 Gilman Drive, La Jolla, CA 92093, USA\\
$^{6}$Department of Astronomy, Box 351580 University of Washington, Seattle, WA 98195\\
$^{7}$National Radio Astronomy Observatory, 520 Edgemont Road, Charlottesville, VA 22903-2475, USA\\
$^{8}$Max-Planck-Institut f\"{u}r extraterrestrische Physik, Giessenbachstra\ss e 1, D-85748 Garching, Germany\\
$^{9}$University of Wisconsin, Department of Astronomy, 475 N Charter St., Madison, WI 53706, USA \\
$^{10}$Max Planck Institut f{\"u}r Astronomie, K{\"o}nigstuhl 17, D-69117 Heidelberg, Germany}

\date{Draft date: \today}

\pubyear{2021}

\begin{document}
\label{firstpage}
\pagerange{\pageref{firstpage}--\pageref{lastpage}}

\maketitle

\begin{abstract}
Previous work has argued that atomic gas mass estimates of galaxies from 21-cm \hi emission are systematically low due to a cold opaque atomic gas component.
If true, this opaque component necessitates a $\sim35\%$ correction factor relative to the mass from assuming optically-thin \hi emission.
These mass corrections are based on fitting \hi spectra with a single opaque component model that produces a distinct ``top-hat'' shaped line profile.
Here, we investigate this issue using deep, high spectral resolution \hi VLA observations of M31 and M33 to test if these top-hat profiles are instead superpositions of multiple \hi components along the line-of-sight.
We fit both models and find that $>80\%$ of the spectra strongly prefer a multi-component Gaussian model while $<2\%$ prefer the single opacity-corrected component model.
This strong preference for multiple components argues against previous findings of lines-of-sight dominated by only cold \hi.
Our findings are enabled by the improved spectral resolution ($0.42$~\kms), whereas coarser spectral resolution blends multiple components together.
We also show that the inferred opaque atomic ISM mass strongly depends on the goodness-of-fit definition and is highly uncertain when the inferred spin temperature has a large uncertainty.
Finally, we find that the relation of the \hi surface density with the dust surface density and extinction has significantly more scatter when the inferred \hi opacity correction is applied.
These variations are difficult to explain without additionally requiring large variations in the dust properties.
Based on these findings, we suggest that the opaque \hi mass is best constrained by \hi absorption studies.
\end{abstract}


\section{Introduction}
\label{sec:intro}

The fundamental difficulty in constraining the atomic interstellar medium (ISM) gas mass is the unknown temperature and density structure of the atomic gas along the line-of-sight.
The temperature and density of atomic hydrogen (\hi) are expected to vary by a factor of $\sim100$ between the warm neutral medium (WNM) and cold neutral medium (CNM) \citep{Field1969ApJ...155L.149F,Wolfire2003ApJ...587..278W}, with an unstable intermediate state between these two phases \citep{Heiles2003ApJ...586.1067H} that accounts for $<20\%$ of the gas \citep[][]{MurrayStanimirovic2018ApJS..238...14M,HillMacLow2018ApJ...862...55H}.
There are consequences on the resulting 21-cm \hi spectra due to variation of the line opacity in these different states. The WNM is primarily optically thin, while the CNM can become optically thick at higher densities.
However, the relative importance of these two components can be difficult to assess.

The CNM can be unambiguously detected in absorption against a bright background source \citep[e.g.,][]{MurrayStanimirovic2018ApJS..238...14M,JamesonMcClure-Griffiths2019ApJS..244....7J} which allows the spin temperature and opacity to be measured.
By comparing to the nearby \hi emission, the fraction of CNM, WNM and intermediate phases can then be estimated.
However, there is limited spatial coverage for lines-of-sight towards sufficiently bright background sources to measure the \hi absorption\footnote{Additionally, \hi self-absorption, where CNM lies in the foreground of bright WNM emission, can provide better spatial coverage where this geometry along a line-of-sight is available \citep{Gibson2005ApJ...626..195G}.}, making it difficult to constrain the total mass in each atomic ISM phase.
These absorption techniques are most useful for Milky Way observations where studies can choose to focus on the brightest quasars as background sources since \hi is detected along most lines-of-sight from our location in the Galactic disc \citep{Heiles2003ApJ...586.1067H,MurrayStanimirovic2018ApJS..238...14M,WangBihr2020A&A...634A.139W}.
Furthermore, the comparison between \hi absorption and emission to constrain the atomic ISM phases requires similar angular resolution in both observations, which is easiest to achieve in the Milky Way but becomes difficult in external galaxies.
In contrast, spatially contiguous \hi emission can easily be traced but it is difficult to separate the atomic ISM phases from the emission alone due to overlapping spectral components and turbulent line broadening.

Extragalactic \hi observations typically resolve $>100$~pc scales, resulting in observed spectra that trace a mixture of unresolved WNM and CNM components.
The WNM contributes a larger fraction of the atomic mass and has a larger volume filling factor relative to the CNM \citep{Ferriere2001RvMP...73.1031F,MurrayStanimirovic2018ApJS..238...14M}, and thus the extragalactic \hi emission is usually assumed to be optically thin.
This assumption is convenient for measuring the atomic ISM mass as the optically thin \hi integrated intensity is (i) independent of the spin temperature and (ii) proportional to the column density.
Since the contribution from opaque CNM emission is not accounted for, using the optically thin assumption will underestimate the atomic ISM mass.
However, when the CNM structures are unresolved on these large scales, it is difficult to estimate what the missing opaque \hi mass fraction is.

One potential solution to the unknown \hi opacity is to explore \hi emission spectra on $<100$~pc scales where the largest CNM structures may be resolved (similar to giant molecular cloud scales) as observed in \hi self-absorption in the Milky Way \citep[][]{Gibson2005ApJ...626..195G,WangBihr2020A&A...634A.139W}.
If \hi emission is resolved (or at least fills the telescope beam) but is opaque, the spectrum is no longer Gaussian and instead is predicted to have a ``top-hat'' shape, where the observed brightness temperature saturates at the \hi spin temperature\footnote{For the simplest case of a single \tspin along the line-of-sight.} \tspin when the emission becomes optically thick \citep{RohlfsBraunsfurth1972AJ.....77..711R}.
Observing such features provides a potentially powerful method to identify the opaque CNM since the opaque ``top-hat'' shape saturates at \tspin and constrains the opacity $\tau$.

With current radio interferometers, resolving $<100$~pc scales limits \hi observations to the nearest galaxies, in particular, those in the Local Group ($D<1$~Mpc).
\citet{Braun2009ApJ...695..937B} explored this approach in M31 with $\sim100$~pc \hi observations, where they noted the presence of ``top-hat'' shaped spectra and modelled these spectra as arising from a single cold, opaque \hi component.
They further advocated that these features are arranged into $\sim100$~pc cold \hi clouds.
\citet{Braun2012ApJ...749...87B} extended this modeling to \hi observations of the LMC and M33 \citep{KimStaveley-Smith2003ApJS..148..473K,Gratier2010A&A...522A...3G}.
Together, they estimated that correcting for the \hi opacity accounts for an additional $\sim35\%$ of \hi mass relative to the mass from the optically thin assumption.
\citet{Braun2012ApJ...749...87B} then used this correction factor to constrain the opaque corrected \hi distribution function with redshift $z=0$.

An implicit assumption in the \citet{Braun2009ApJ...695..937B} and \citet{Braun2012ApJ...749...87B} modeling is that most lines-of-sight through these galaxies are dominated by a single \hi component, and further, that that component is primarily cold ($<100$~K) \hi.
This differs from the structure typically observed in the Milky Way and Magellanic Clouds, where few \hi studies find ``top-hat'' \hi components \citep{LeeStanimirovic2015ApJ...809...56L,JamesonMcClure-Griffiths2019ApJS..244....7J}.
Of particular note is the lack of top-hat spectra from \citet{StanimirovicMurray2014ApJ...793..132S} on $\sim1$~pc scales towards the Perseus molecular clouds.
Since the \citet{Braun2009ApJ...695..937B} model identifies beam-filling opaque emission, the lack of such features in Perseus on much smaller scales raises the question of whether these conclusions are driven by the analysis approach or are unique to the environments observed.

Comparisons with other tracers of the neutral ISM column density also show different trends compared to the \hi opacity-corrected column density from the \citet{Braun2009ApJ...695..937B} method.
For example, molecular clouds traced by bright CO emission have the highest neutral ISM column density and thus are likely places where opaque \hi should be detected.
However, there is not a strong relation between CO emission and the opaque \hi regions from \citet[][their figures 11, 12, \& 16]{Braun2009ApJ...695..937B}, in contrast to what is seen in the Milky Way, where the largest \hi self-absorption structures are strongly correlated with CO emission and molecular gas \citep{Gibson2005ApJ...626..195G,WangBihr2020A&A...634A.139W}.

Studies that do find top-hat \hi components note the difficulty in distinguishing from two overlapping velocity components \citep{RohlfsBraunsfurth1972AJ.....77..711R}.
Some studies use spatial continuity to confirm such features are indeed from $>1$ \hi components overlapping in velocity \citep{PeekHeiles2011ApJ...735..129P}.
One limitation of the \citet{Braun2009ApJ...695..937B} and \citet{Braun2012ApJ...749...87B} studies is the $1.4{-}2.3$~\kms velocity resolution of their observations.
Assuming a spectrally resolved Gaussian component requires sampling with channels $\sim2\times$ the RMS line width \citep{KochRosolowsky2018RNAAS...2..220K}, and thus the minimum thermal temperatures for a resolved component with these channel widths are $1300{-}3600$~K, far larger than the CNM temperatures.
Our initial work with these new observations in M33 \citep{Koch2018MNRAS} demonstrated that a finer spectral resolution can spectrally resolve features that may not have been evident in previous observations.

Given these open questions, we investigate the topic of \hi opacity using  new \hi observations of M31 and M33 with $0.42$~\kms spectral resolution taken with the Karl G. Jansky Very Large Array (VLA). In particular, we address the question: do the observed \hi line shapes in M31 and M33 imply a single opaque \hi component, as suggested by \citet{Braun2009ApJ...695..937B}, or are the spectra better described by multiple components along the line-of-sight?
Answering this question has broad implications for our understanding of the neutral ISM in the nearby universe, since the ISM mass is dominated by the atomic phase \citep{SaintongeCatinella2017ApJS..233...22S,WalterCarilli2020arXiv200911126W}, and constraints on the cosmic density of \hi in galaxies at $z=0$ depend on our ability to accurately measure the atomic gas mass.
In this work, the key observational improvement is the fine spectral resolution that can resolve the $0.85$~\kms thermal line width of 120~K cold \hi gas while spatially resolving $60{-}300$~pc (accounting for galaxy inclination), closely matching the spatial scales used in previous studies.

We introduce the new \hi observations in \S\ref{sec:observations} and the model fitting and metric for comparison in \S\ref{sec:hi_models}.
In \S\ref{sec:mod_comp}, we use the model comparison and present our main result: most \hi spectra in M31 and M33 are best-modelled with a multi-component Gaussian model.
We then compare the atomic surface density to tracers of dust in \S\ref{sec:dustcomparisons}, where we find that including the inferred opaque \hi correction would require large variations in dust properties to explain.
From these results, we discuss in \S\ref{sec:discussion} the discrepancies with previous work, the uncertainty on the opacity-corrected atomic ISM mass, comparisons with \hi absorption studies, and the future outlook on using \hi emission surveys to identify cold atomic ISM.
Finally, we summarize our conclusions in \S\ref{sec:summary}.

Throughout this work, we adopt distances of $744$~kpc for M31 \citep{Vilardell2010A&A...509A..70V} and $840$~kpc for M33 \citep{Freedman2001ApJ...553...47F}.

\section{Observations}
\label{sec:observations}

\subsection{VLA and Single-dish \texorpdfstring{\hibold}{HI} observations} 
\label{sub:vla}

We obtained new \hi observations for M31 and utilize our recent \hi data set for M33 obtained by the VLA.
Table~\ref{tab:observe_comp} presents basic properties of these VLA observations and those used in \citet{Braun2009ApJ...695..937B} and \citet{Braun2012ApJ...749...87B} for comparison.
The key property of our new observations is the fine spectral resolution ($0.42$~\kms), which is a factor of $3{-}5$ finer than in previous observations.

The M33 observations are presented and described in detail in \citet{Koch2018MNRAS}; we briefly summarize the observations here.
We observed a 13-pointing mosaic that covers the inner $\sim12$~kpc of M33 in C-configuration (project 14B-088) for a total of 52~hr.
The nearby source 3C48 was used as the flux, delay, and gain calibrator.
The \hi line was observed with a 977~Hz resolution, corresponding to a spectral channel width of $0.206$~\kms, and the naturally-weighted beam size is $19\arcsec$ ($\sim80$~pc).
These VLA observations were combined with existing GBT data to provide short-spacing information \citep[see ][]{Lockman2012AJ....144...52L}.
To match the M31 spectral resolution of $0.42$~\kms (see below), we spectrally downsample the $19\arcsec$ M33 \hi cube from the original $0.206$~\kms from \citet{Koch2018MNRAS}.
The typical \hi spectral widths identified in \citet{Koch2018MNRAS} and \citet{KochRosolowsky2019MNRAS.485.2324K} are larger than $0.42$~\kms and so we do not expect downsampling the spectral resolution by a factor of 2 will affect measurements of the spectral shapes.

The new M31 observations cover a 7-point mosaic in the northern half of the galaxy that overlaps with the \textit{Hubble Space Telescope} observations from the PHAT survey \citep{DalcantonWilliams2012ApJS..200...18D} and the CARMA \coonezero survey field \citep[A.~Schruba et al., in preparation; for a first presentation see][]{Caldu2016AJ....151...34C}.
We combined observations taken in the VLA's B, C, and D-configurations\footnote{The D-configuration observations include a full mosaic of M31 that will be presented in E.~W.\ Koch et al.\ (in preparation)} (projects 14A-235 and 15A-175) and use a $uv$-taper when imaging to produce a mosaic with $18\arcsec$ beam size, roughly matching the C-configuration VLA mosaic of M33.
The \hi line was observed with $1.95$~kHz spectral channels, corresponding to a velocity resolution of $0.42$~\kms.
We reduced the data using the standard VLA pipeline with CASA v5.4.1, with minor modifications for spectral line data, including: (i) splitting the continuum and line spectral windows; (ii) disabling Hanning smoothing to retain the $0.42$~\kms velocity resolution for the \hi data; and (iii) avoiding RFI flagging and re-weighting (\textsc{statwt}) over M31's velocity range.
Due to the overlap between the red-shifted side of M31 and Milky Way \hi emission, we also flagged velocity ranges with \hi absorption towards the bandpass calibrator and interpolated across this gap.
For each track, we iteratively re-ran the pipeline after checking the data and applying (typically) minor manual flagging.
Most tracks required 2 pipeline runs to produce well-calibrated data.

We imaged the M31 \hi data in $0.42$~\kms channels using a similar approach to that used for the M33 imaging described in \citet{Koch2018MNRAS}.
To include short-spacing information, we combined the VLA \hi cube with the single-dish data from the Effelsberg-Bonn \hi Survey \citep[EBHIS;][]{WinkelKerp2016A&A...585A..41W}.
In Appendix~\ref{app:imaging_approach}, we describe the imaging approach and combination with the EBHIS data in detail.

\begin{figure*}
\includegraphics[width=\textwidth]{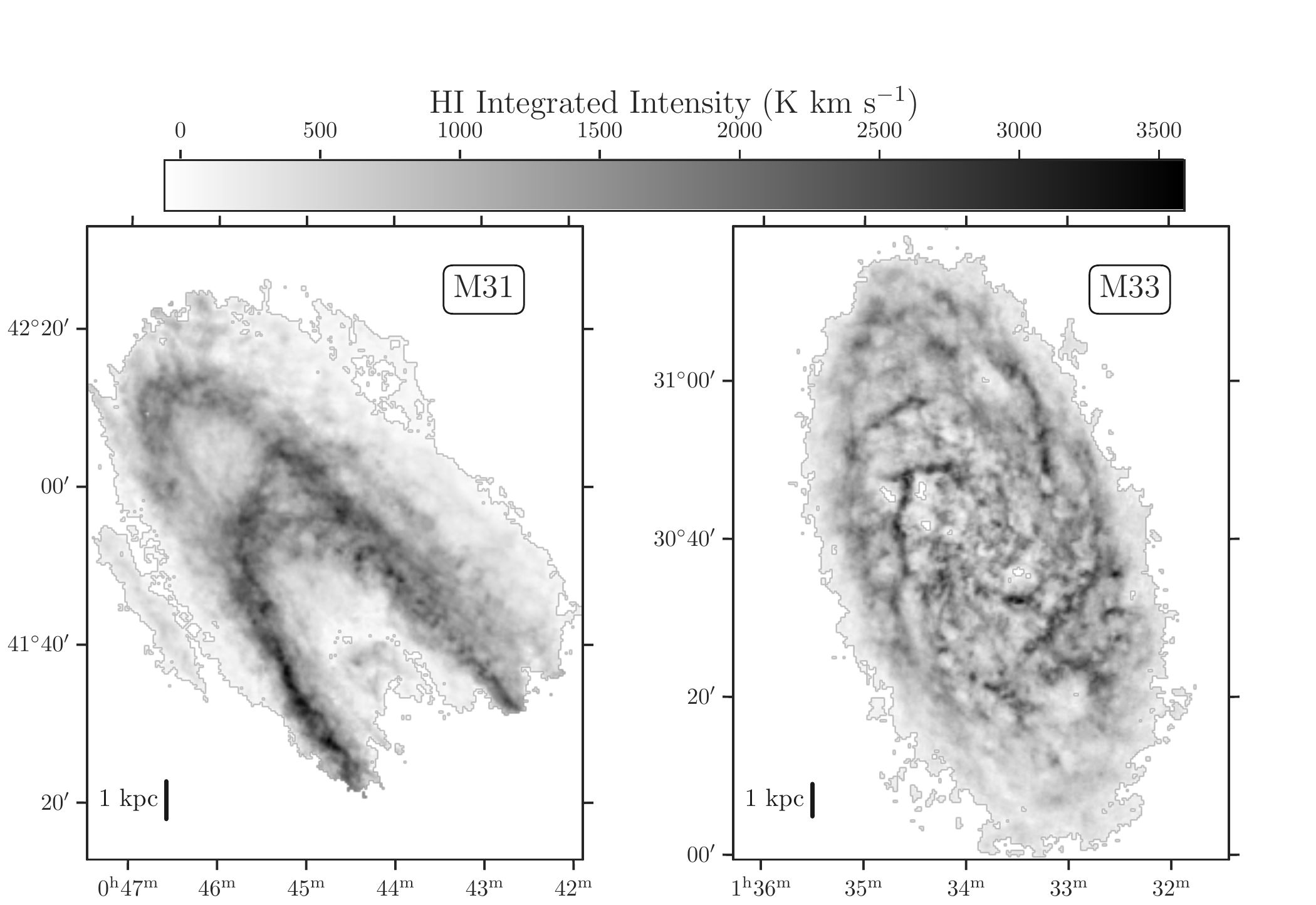}
\caption{\label{fig:m31_m33_maps}
\hi integrated intensity maps of M31 and M33 (see Table~\ref{tab:observe_comp}).
Our new VLA \hi map of M31 covers most of the Northern half of M31.
The VLA \hi map of M33 is presented and described in detail in \citet{Koch2018MNRAS}.
}
\end{figure*}

Figure~\ref{fig:m31_m33_maps} shows the \hi integrated intensity maps for M31 and M33 at the resolution used in this work.
The remainder of this section describes additional data handling required for this work.

For the M31 map, the bright background radio source at ICRS $0^{\rm h}46^{\rm m}48^{\rm s} \ {+}42\degree08\arcmin56\arcsec$ falls within the field-of-view leading to strong \hi absorption features from M31 \citep[labelled as Source \#497 in the catalogue from][]{Braun1990ApJS...72..761B}.
These features persist after subtracting the continuum emission from the $uv$ data prior to imaging. 
We mask the region around this source by defining a region $1.5\times$ the beam area to avoid fitting strong absorption features, for which the methods in \S\ref{sec:hi_models} are not suited.

The per-channel rms noise is $2.8$~K for M31 and $2.0$~K for M33 in the spectrally-matched data.
Since strong emission continues to the edges of the observed area, particularly for M31 (Figure~\ref{fig:m31_m33_maps}), the noise varies with the effective primary beam for the mosaic.
We weight the noise levels used for the spectral fitting in \S\ref{sec:hi_models} by the primary beam, which increases the noise levels by up to $\sim5\times$ at the edges of the mosaic.
Since the M31 mosaic only covers a part of the galaxy, \hi signal is detected in some places along the edge of the mosaic and is affected by this increase in the noise.

Throughout the paper we use the M31 and M33 \hi cubes at their respective angular resolutions.
Because of the differences in distance and inclination between the two galaxies, the linear physical resolution overlaps at the respective original angular resolutions.
The linear resolution for M31 is $60\times300$~pc at $18\arcsec$ along the major and minor axes, respectively.
This large range is due to M31's large inclination angle of $i=78\degree$ \citep[][]{Corbelli2010A&A...511A..89C} that tilts the plane of sky in the observed frame.
For M33, the linear resolution ranges from $80\times130$~pc, which is smaller due to the lower inclination angle of $i=55.1\degree$ \citep{Koch2018MNRAS}.
We tested whether matching the angular resolution affects the model fits from \S\ref{sec:hi_models} by fitting small regions of M31 and M33 at matched angular resolution.
We found that the fits were consistent at the matched and original angular resolutions.

To convert the \hi integrated intensity to atomic gas surface density, we use the optically thin factor $0.0196$ \msolpcsq (\kks)$^{-1}$ which includes a $1.36$ factor for helium and metals.

\begin{table*}
    \centering
    \begin{tabular}{ccllllll}
Reference & Target & Resolution ($\arcsec$) & Linear resolution (pc) & Channel  & Noise per channel & $5\sigma_{\rm rms}$ optically-thin & Observatory \\
                  & & & Major $\times$ Minor axis & Width (\kms) & ($\sigma_{\rm rms}$; K) & column density (cm$^{-2}$) & \\\hline
\citet{Braun2009ApJ...695..937B} & M31    & 30                     & $100\times500$    & 2.3      & 1.0  &  $2.0\times10^{19}$    & WSRT     \\
This Work                        & M31    & 18                     & $60\times300$     & 0.42      & 2.8  &  $9.8\times10^{18}$    & VLA    \\[6px]
\citet{Braun2012ApJ...749...87B} & M33    & 30                     & $130\times205$    & 1.4      & 2.1  &  $2.6\times10^{19}$    & VLA  \\
This Work                        & M33    & 19                     & $80\times130$     & 0.42      & 2.0  &  $7.0\times10^{18}$    & VLA      
\end{tabular}

    \caption{Summary of 21-cm \hi observations from this work and those from \citet{Braun2009ApJ...695..937B} and \citet[][see their Table 1]{Braun2012ApJ...749...87B}.
    The M33 VLA observations used in \citet{Braun2012ApJ...749...87B} are presented in \citet{Gratier2010A&A...522A...3G} \citep[see also][]{Thilker2002ASPC..276..370T}.
    The linear resolution range accounts for the galaxy inclination.}
    \label{tab:observe_comp}
\end{table*}

\subsection{Dust emission and extinction}
\label{sub:dustmaps}

Dust emission and dust extinction provide independent measurements of ISM column density since it is expected to be well-mixed with gas on the $>60$~pc scales we explore here.
Here, we use both dust emission and extinction maps to compare with the gas column density inferred from \hi.

First, we use the dust surface density maps of M31 and M33 from \citet{UtomoChiang2019ApJ...874..141U} with $167$~pc resolution in the plane of sky.
These surface density maps are derived from a modified blackbody model fit to the spectral energy distribution (SED) in 5 bands from the \textit{Herschel Space Observatory} from $100{-}500$~$\mu$m using a broken power-law emissivity model, following the approach of \citet{Gordon2014ApJ...797...85G} and \citet{Chiang2018ApJ...865..117C}.

Second, we use the dust extinction $A_{\rm V}$ map of M31 over the region observed by PHAT from \citet{DalcantonFouesneau2015ApJ...814....3D}.
In this map, the dust extinction is modelled within a grid of 25~pc ($6.5\arcsec$) square regions by fitting the SEDs of $\sim10$s of stars to a log-normal distribution in extinction $A_{\rm V}$.
From this modeling, we use the map of mean $A_{\rm V}$ values to compare with other tracers of the ISM column density.

\section{Modelling \texorpdfstring{\hibold}{HI} spectra} 
\label{sec:hi_models}

The VLA \hi observations we present have a similar sensitivity and spatial resolution as previous M31 and M33 observations.
They differ, however, in their $3{-}5$ times finer spectral resolution ($0.42$~\kms), which reveals ubiquitous spectral complexity in the \hi spectra, as noted in \citet{Koch2018MNRAS} for M33.

\begin{figure*}
\includegraphics[width=\textwidth]{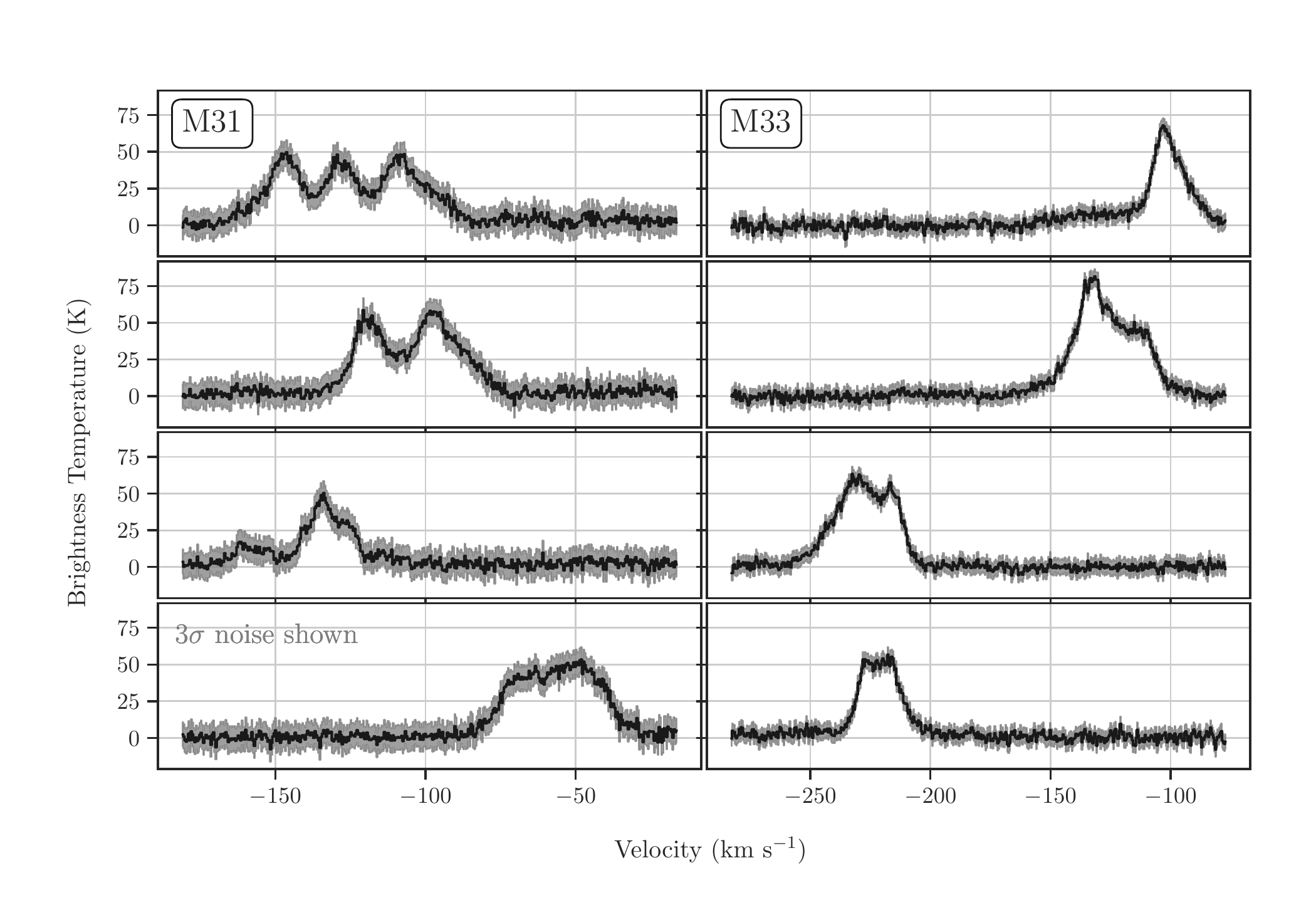}
\vspace*{-20px}
\caption{\label{fig:m31_m33_spectra}
Example \hi spectra in M31 (left) and M33 (right).
The spectral channel width in all spectra is $0.42$~\kms, and the $3\sigma$ rms noise levels are shown in gray to emphasize that the spectral complexity is real.
In both galaxies, we find distinct spectral features that consistently deviate from a single Gaussian profile.
Most of these example spectra have multiple peaks, though notably, the spectra in the bottom row show the general ``flat-top'' shape that may result from opaque \hi.
By consistently modeling the \hi spectra with the multi-Gaussian and opaque models, we quantitatively distinguish between the appropriate model for the array of spectral shapes.
}
\end{figure*}

Figure~\ref{fig:m31_m33_spectra} shows 8 spectra---4 from each galaxy---that demonstrate a range of \hi line profiles.
These different profiles present an interesting opportunity to compare models for \hi spectra.

In this section, we explicitly test two models to describe the \hi line shape: a single opaque \hi component model, as advocated by \citet{Braun2009ApJ...695..937B} and \citet{Braun2012ApJ...749...87B}, and a multi-component Gaussian model, that implicitly assumes that profiles are due a superposition of optically thin clouds along a line-of-sight.
We first introduce these models, the fitting procedures, and the model selection used to compare between them.

\subsection{Opaque \texorpdfstring{\hibold}{HI} model}
\label{sub:optthick_model}

\begin{figure}
\includegraphics[width=0.5\textwidth]{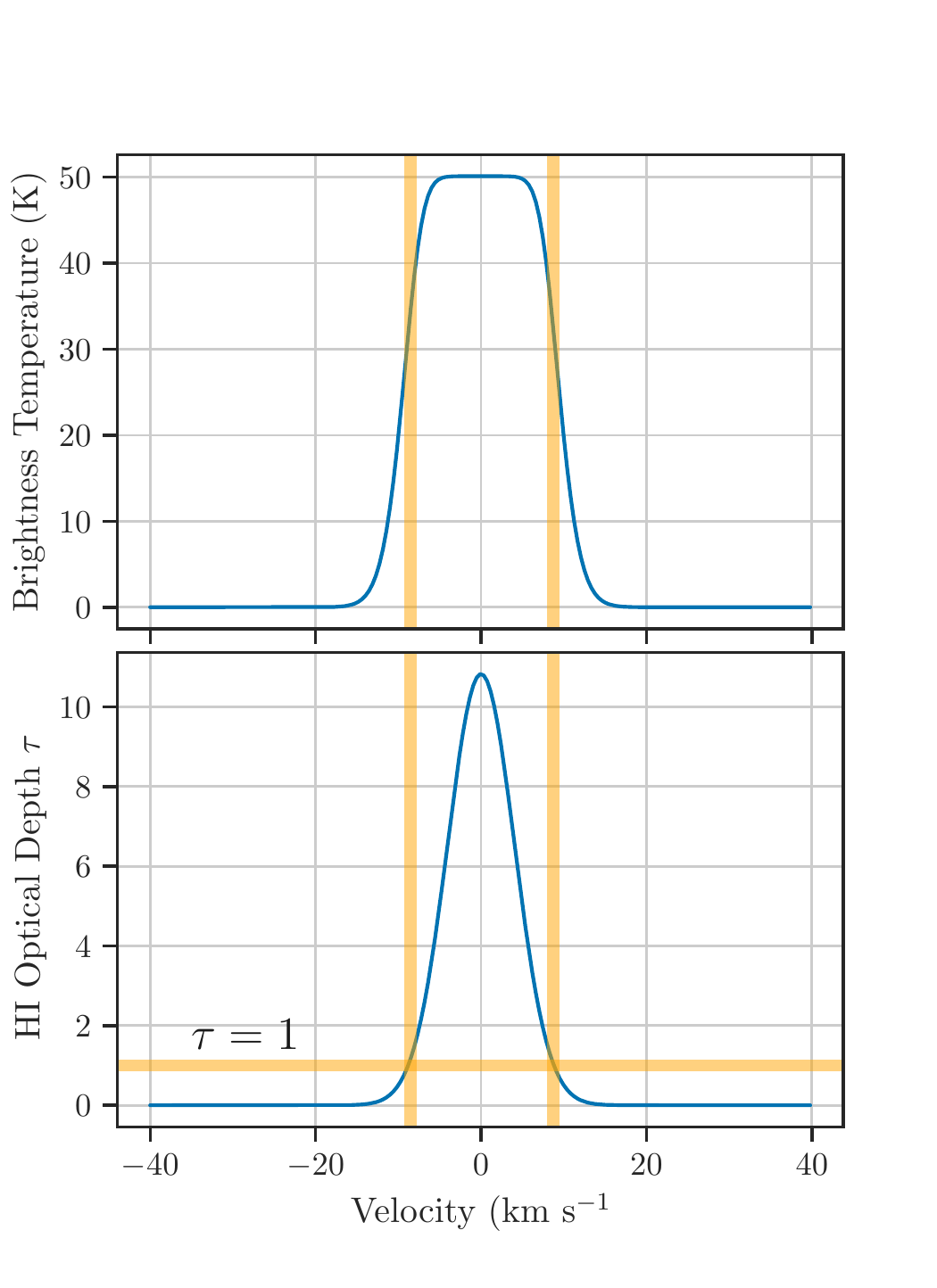}
\vspace*{-30px}
\caption{\label{fig:opaque_model}
A single opaque \hi component model (Equation~\eqref{eq:thickhi_profile}; top) where the \hi line opacity follows a Gaussian profile (Equation~\eqref{eq:tau_profile_simple}; bottom).
This example uses a peak temperature of $T_{\rm p}=540$~K, a spin temperature of \tspin$=50$~K, and a line width of $\sigma=4$~\kms.
The corresponding column density for this model is $N_{\rm HI}=10^{22}$~\cmtwo ($\Sigma_{\rm HI}=80$~\msolpcsq).
The orange vertical lines show where the emission becomes opaque at $\tau=1$.
Within the opaque velocity region, the observed line profile has a distinct ``top-hat'' shape that saturates at \tspin.
If the emission were to be treated as optically thin ($\propto$~integrated intensity), the \hi column density, and therefore mass, would be significantly underestimated.
}
\end{figure}

In the absence of a background continuum source, the general line profile for a single component of isothermal \hi broadened by turbulence is
\begin{equation}
    \label{eq:thickhi_profile}
    T_{\rm b}(v) = T_{\rm s} \big[ 1 - {\rm exp}\left[-\tau(v) \right] \big],
\end{equation}
where $v$ is the velocity\footnote{Or an equivalent spectral quantity.}, $T_{\rm b}$ is the observed brightness temperature, $T_{\rm s}$ is the \hi spin temperature, and $\tau(v)$ is the optical depth profile.
We also note that this assumes a diffuse continuum component is negligible \citep[see eq. 11 \& 12 in][]{LeeStanimirovic2015ApJ...809...56L}.
Following \citet{Braun2009ApJ...695..937B}, we assume that the optical depth profile is a Gaussian with thermal and turbulent line broadening.
The optical depth profile can then be expressed in terms of the \hi column density ($N_{\rm HI}$; \cmtwo), \tspin (K), and line width ($\sigma$; \kms) as
\begin{equation}
    \label{eq:tau_profile}
    \tau(v) = \frac{5.49\times10^{-19} N_{\rm HI}}{\sqrt{2 \pi} T_{\rm s} \sigma} \ {\rm exp} \left[- \frac{(v - v_{\rm p})^2}{2\sigma^2} \right],
\end{equation}
where $v_{\rm p}$ is the velocity at the peak in optical depth.
The line width can be expressed in terms of a thermal and non-thermal component, $\sigma = \sqrt{\sigma_{\rm t}^2 + \sigma_{\rm nt}^2}$, where the non-thermal component is presumed to be dominated by turbulence.
Similar to \citet{Braun2009ApJ...695..937B}, we assume the kinetic and spin temperatures are equal ($T_{\rm k}=T_{\rm s}$) in the absence of additional constraints, which is correct for the CNM only.
In the limit of high \tspin, high $\sigma$, or low $N_{\rm HI}$, we recover the optically thin limit of a Gaussian line shape since ${\rm exp}\left[-\tau \right]\approx1-\tau$ for small $\tau$.
Figure~\ref{fig:opaque_model} shows an example of the model shape from Equations~\eqref{eq:thickhi_profile} and~\eqref{eq:tau_profile} where the opacity significantly affects the observed line shape that produces a ``top-hat'' shape.

We stress that Equation~\eqref{eq:thickhi_profile} describes the line profile for a single component of \hi, though it is common in, for example, Milky Way \hi observations for many components to be detected along a line-of-sight.
The issue in extending this opacity-corrected \hi model to multiple components is the strong degeneracy between the fit parameters.
In particular, the ``top-hat'' shape that constrains the line opacity must be spectrally resolved from other components, as blended structures can be fit by both a top-hat shape and a combination of Gaussians (see Figure~\ref{fig:opaque_model}).
For this reason, previous studies typically fit only a single component to each spectrum \citep{Braun2009ApJ...695..937B,Braun2012ApJ...749...87B}, or at most two components \citep{PeekHeiles2011ApJ...735..129P}.
Here, we also fit only with a single opacity-corrected component to compare with \citet{Braun2009ApJ...695..937B} and \citet{Braun2012ApJ...749...87B}.

While Equation \eqref{eq:thickhi_profile} assumes a single isothermal component, we expect that the \hi measured within the beam will be a mixture of both WNM and CNM.
For the \hi opacity to be large enough to observe a top-hat shaped spectrum, the CNM must dominate by mass within the telescope beam.
In our observations and those in \citet{Braun2012ApJ...749...87B}, observing a top-hat shaped spectrum implies that cold atomic ``clouds'' dominate the atomic ISM mass on $\sim100$~pc scales.
If the WNM dominates instead, the observed spectrum will converge to the optically thin case, as is often assumed in studies of nearby galaxies with coarse spatial resolution.

\citet{Braun2009ApJ...695..937B} and \citet{Braun2012ApJ...749...87B} fit \hi spectra in M31, M33, and the LMC with $\tau$ as defined in Equation~\eqref{eq:tau_profile}.
They justified using an isothermal approximation to describe the \hi based on the low filling fraction of cold \hi clouds found by \citet{Braun1997ApJ...484..637B}.
This low filling fraction is inferred from the ``high-brightness network'' of \hi where \citet{Braun1997ApJ...484..637B} suggest the \hi is most likely to be opaque.
\citet{Braun2009ApJ...695..937B} fit \hi spectra by defining a grid of pre-computed models and minimizing the reduced $\chi^2$ to identify the best-fit model.
To limit the range of possible models in the grid, the centre of the line was fixed to within $\pm5$~\kms of the velocity centroid, or about $\pm2$~channels in their observations.
To summarize, the fitting method from \citet{Braun2009ApJ...695..937B} has 4 free parameters: \tspin, $N_{\rm HI}$, $\sigma_{\rm nt}$, and $v_{\rm p}$.

Because 3 of the 4 free parameters contribute to the first term in the optical depth (Equation~\eqref{eq:tau_profile}), these parameters are strongly covariant in the fit.
To measure and reduce the parameter covariance from from Equations~\eqref{eq:thickhi_profile} and \eqref{eq:tau_profile}, we make three changes to the fitting approach from \citet{Braun2009ApJ...695..937B}:

\begin{enumerate}
    \item We fit the model using a Levenberg-Marquardt algorithm, as implemented in the \textsc{scipy} python package \citep{2020SciPy-NMeth}, instead of a grid search method.
    We then estimate the parameter uncertainty from the covariance matrix of the fit.
    Parameter covariances become important as the model approaches the optically thin limit, where $\sigma$, $N_{\rm HI}$, and $T_{\rm s}$ will be strongly covariant due to the lack of additional constraints on $T_{\rm s}$.
    
    \item To minimize parameter covariance, we do not split the line width into a thermal and non-thermal component.
    This change is important when the profile approaches the optically thin limit and \tspin is unconstrained, which leads to the thermal line width contribution being limited only by the upper limit of \tspin.
    Here we adopt an upper limit of $T_{\rm s}=8000$~K for \hi to match the range\footnote{We note that $T_{\rm s}=8000$~K falls within the WNM range found by \citet{MurrayLindner2014ApJ...781L..41M} from deep \hi absorption observations in the Milky Way.} used by \citet{Braun2009ApJ...695..937B}.
    When \tspin is constrained, the non-thermal line width can be recovered by subtracting the thermal component in quadrature.

    \item We further decrease the parameter covariance in the model by simplifying the optical depth profile from Equation~\eqref{eq:tau_profile}.
    We define a \tpeak peak temperature such that\footnote{For reference, the parameter space that \citet{Braun2009ApJ...695..937B} define allows for $T_{\rm p, max}=3.47\times10^4$~K, where $N_{\rm HI, max}=3.1\times10^{23}$~\cmtwo and $\sigma_{\rm min}=2.0$~\kms. We set an upper limit of $T_{\rm p, max} \leq 5 \, {\rm max}\left( T_{\rm B} \right)$ for each spectrum; raising this limit did not affect our results.}, $T_{\rm p}= 5.49\times10^{-19} N_{\rm HI} / \sqrt{2\pi} \sigma$.
    Then the peak optical depth is simplified to be $\tau_{\rm p}=T_{\rm p} / T_{\rm s}$ and Equation~\eqref{eq:tau_profile} becomes:
    \begin{equation}
        \label{eq:tau_profile_simple}
        \tau(v) = \left( \frac{T_{\rm p}}{T_{\rm s}} \right) {\rm exp} \left[ - \frac{(v - v_{\rm p})^2}{2\sigma^2} \right].
    \end{equation}
    For opaque \hi, the observed peak temperature will converge to \tspin, while for optically thin emission it will converge to \tpeak.
\end{enumerate}

Using this change of variables removes the added covariance between \tspin and $\sigma$ from the thermal and non-thermal line width components, and introduces a parameter \tpeak with a similar range to \tspin, replacing $N_{\rm HI}$ which varies over several orders of magnitude.
The line opacity is then primarily set by the peak optical depth, $\tau_{\rm p} = T_{\rm s} / T_{\rm p}$.

We keep similar constraints on the velocity of the line centre.
Due to the finer spectral resolution of our data, a limit of $\pm5$~\kms translates to $\pm12$~channels.

\subsection{Multi-Gaussian model}
\label{sub:gauss_model}

We also fit the observed \hi spectrum to a multi-component model of Gaussians.
Compared to the opaque \hi model above, this multi-component Gaussian model allows for multiple components along a line-of-sight, but restricts each component to a Gaussian shape, i.e., being optically thin.
As we described below, overlapping Gaussians become degenerate and adding additional free parameters to account for opacity makes the fitting intractable, or at best, poorly constrained due to the strong covariance in parameters.

The general form for a multi-component model comprised of $N$ Gaussians is:
\begin{equation}
    \label{eq:multigauss}
    T_{\rm b}(v) = \sum_i^N T_{{\rm p}, i}(v) = \sum_i^N T_{{\rm p}, i} \ \textrm{exp}\left[- \frac{(v - v_{0, i})^2}{2\sigma_{i}^2} \right],
\end{equation}
where each component is defined in terms of its peak temperature ($T_{{\rm p}, i}$), central velocity ($v_{0, i}$), and line width ($\sigma_{i}$).

\citet{RohlfsBraunsfurth1972AJ.....77..711R} first point out that Milky Way \hi spectra with flat-tops could also be explained by two or more Gaussian components blended together.
The difficulty in distinguishing between a single optically thick component and multiple optically thin Gaussian components lies in the flexibility of a multi-Gaussian model: two or more Gaussians can be arranged to explain many spectral shapes and do not form an orthogonal basis set.
In this section, we describe our approach for determining an appropriate number of Gaussian components to avoid overfitting, while accounting for spatial continuity to distinguish between one or more overlapping Gaussians.

Several recent studies present algorithmic approaches for \mbox{(semi-)}\linebreak[0]{}automated spectral line fitting of Gaussian components \citep[e.g.,][]{Haud2000A&A...364...83H,Lindner2015AJ....149..138L,Henshaw:2016el,KeownDiFrancesco2019ApJ...885...32K,MarchalMiville-Deschenes2019A&A...626A.101M,SokolovPineda2020ApJ...892L..32S}.
The method we use here combines elements from some of these studies to optimize the fitting for our particular data sets, though we highlight that we only include spatial information from nearest neighbours rather than fitting all neighbouring spectra together (\S\ref{subsub:final_fit}).
From our early testing of different methods, we note that most spectral line fitting algorithms will likely require at least small changes to be optimized for a particular data set or science goal.
Fortunately, the publicly available codes from these projects enable specific optimization with relative ease.

Our multi-Gaussian fitting method has three stages: (1) identifying the number of components and initial parameter guesses; (2) an initial non-least squares optimization and internal model selection test; and (3) a nearest neighbour model selection test and final fit.
Each stage is described in the following sections.

\subsubsection{Stage 1: Number of components and parameter estimates}
\label{subsub:component_guesser}

The biggest issue in fitting a multi-Gaussian model is choosing the appropriate number of components to fit the spectrum with.
As mentioned above, part of this difficulty is the correlation between different components.
To overcome this issue, the number of components and initial estimates for the components can be recovered by using information from the spectrum shape.

Here, we use the ``derivative spectroscopy'' implemented in \textsc{gausspy} \citep{Lindner2015AJ....149..138L}, which we briefly describe below.
Estimates of the derivative from the finite-difference between channels are strongly susceptible to noise.
To overcome this, smoothing can include the information over several channels at each point in the spectrum.
Common approaches for this smoothing include Gaussian or median smoothing \citep{RienerKainulainen2020A&A...633A..14R} or total-variation regularization \citep{Lindner2015AJ....149..138L}.
From this smoothing, derivatives provide information to locate and estimate components.
Peaks in a spectrum are identified based on (i) a negative second derivative minimum, (ii) a zero crossing in the third derivative, and (iii) a positive fourth derivative \citep{Lindner2015AJ....149..138L}.

Previous works using \textsc{gausspy} identified Gaussian components in two steps: first by identifying narrow components with a small smoothing length, followed by identifying wide components with a large smoothing length.
Previous studies using this method identified optimal smoothing lengths using gradient descent from a training set of spectra \citep{Lindner2015AJ....149..138L,MurrayStanimirovic2018ApJS..238...14M,RienerKainulainen2020A&A...633A..14R}.
In this work, we assess the performance of a two-stage identification by-eye on a small set of spectra from our data sets while varying both the narrow and wide smoothing lengths.
Our data contain extremely wide ($\sigma>30$~\kms) features, particularly in M31 where the line-of-sight depth through the disc is large, and we find that the two-stage approach used in previous works does not consistently identify spectral features across this large range in line widths.
Here, we explore a more ``brute-force'' component identification method.

To account for the wide range in spectral shapes in our data, we modify the identification procedure in \textsc{gausspy} to consider a range of smoothing lengths.
Following the smoothing and component identification scheme from \textsc{gausspy}, we identify and estimate the component parameters from small to large smoothing scales.
At each smoothing scale, the component properties are estimated and subtracted from the spectrum.
Then, at progressively larger smoothing scales, wider components are added from the residual in the spectrum.
This procedure closely follows \textsc{gausspy}, just with $>2$ smoothing lengths considered.

\subsubsection{Stage 2: Initial fit and internal model selection}
\label{subsub:initial_fit}

The parameter estimates from the first step are crucial for the multi-component Gaussian fit to converge to a reasonable solution.
The hope is that the initial parameter estimates start the minimization algorithm close enough to the global minimum so that it will converge quickly and not fall into a local minimum that sometimes results from the correlations between Gaussian components.
However, the approximations in step one to get parameter estimates are not always robust and can lead to (i) fits converging to vastly different parameters from the starting point, or (ii) incorrect number of components or their placements.
To account for this, we first fit the spectrum using the estimates from stage one, and then perform an internal model selection to ensure the resulting fit is valid.
Similar to \textsc{gausspy} \citep{Lindner2015AJ....149..138L}, we fit spectra using the Levenberg-Marquardt method implemented in the python package \textsc{lmfit}\footnote{\textsc{lmfit} wraps and extends the optimization algorithm from the \textsc{scipy} library \citep{lmfit}.}.

The model selection used here and in the following sections relies on the Bayesian information criterion (BIC) fit statistic \citep{schwarz1978_bic}.
The BIC is defined as a likelihood plus penalization term, where the latter increases in value with the number of free parameters to avoid overfitting
\begin{equation}
    \label{eq:bic}
    {\rm BIC} = {\rm ln}(m) \ k - 2 \ {\rm ln} (\hat{L})~,
\end{equation}
where $m$ is the number of data points (velocity channels), $k$ is the number of free parameters, and $\hat{L}$ is the likelihood function.
An optimal model fit minimizes the BIC statistic, as the penalization term in Equation~\eqref{eq:bic} is positive.

For the multi-component Gaussian model, we assume the data uncertainties ($\sigma_{\rm rms}$) are independent and normal, and so the log-likelihood function ${\rm ln} (\hat{L})$ for fitting a single spectrum has the standard form of
\begin{equation}
    {\rm ln} (\hat{L}) = - \frac{1}{2\sigma_{\rm rms}^2} \sum_{j=1}^{m} {\big[y_j - T_{\rm b}(v_j)\big]}^2 + C(\sigma_{\rm rms})~,
\end{equation}
where $m$ is the number of velocity channels, and $y_j$ is the observed brightness temperature at velocity $v_j$ and $T_{\rm b}(v_j)$ is the model from Equation~\eqref{eq:multigauss}.
The $C(\sigma_{\rm rms})$ term is a constant that depends only on the noise, which is a constant for all spectral channel and therefore does not change between models.
The preferred model minimizes the BIC.
We choose the BIC since it penalizes additional free parameters more strongly than other common statistics (e.g., the Akaike information criterion).

There are two parts to the internal model selection:
\begin{enumerate}
    \item The integral of each component must exceed $5\sigma$ uncertainty integrated over a resolved Gaussian line.
    This restriction removes spurious narrow features due to noise.
    Following \citet{KochRosolowsky2018RNAAS...2..220K}, we define a resolved line as having $>5$~channels (i.e., $2.1$~\kms) across the full-width-half-maximum (FWHM).
    This restriction requires a minimal integrated intensity of $\sqrt{\pi/(4\ln 2)}\allowbreak (2.1~{\rm km~s^{-1}})\allowbreak (5\sigma_{\rm rms})=31$~K~\kms for M31 and $22$~K~\kms for M33.
    Components below these limits are removed and the spectrum is refit with fewer components.
    
    \item The above procedure is well-suited to identify most spectral features, but components near the noise level may be spurious noise features or may be split into multiple components.
    To ensure the minimum number of components is used in the fit, we next test whether the removal of the smallest integrated intensity substantially changes the ${\rm BIC}$ of the fit.
    Following \citet{kassraftery_bicdiff}, we consider changes of $\Delta {\rm BIC}_{N_i, N_f} > 10$ to indicate a strong preference for the initial model with the initial $N_i$ number of components relative to the simplified model with $N_f$ components.
    Since the test is always with fewer components than the initial fit, $N_i>N_f$.
    We repeat this procedure by iteratively removing components from the model until the ${\rm BIC}$ of the fits converge to $\Delta {\rm BIC}_{N_i, N_f} = {\rm BIC}_{N_i} - {\rm BIC}_{N_f} < 10$.
    The components are removed in order from lowest to highest integrated intensity.

    One key aspect of this procedure is testing the null hypothesis of whether the spectrum is best modelled by \textit{any} Gaussian components.
    We continue the process of iteratively removing components, if needed, until $N_f=0$ 

\end{enumerate}

The updated multi-Gaussian model is fit using the fewest components that passes both selection tests.
In Appendix~\ref{app:gaussian_model_testing}, we explore how well this fitting method recovers Gaussian components from synthetic multi-component Gaussian spectra.

\subsubsection{Stage 3: Nearest neighbour model selection and final fit}
\label{subsub:final_fit}

We introduce one additional step to produce the models used later in this paper.
Because we fit each spectrum independently of its neighbours, the fits do not account for the correlation of nearby pixels and models of neighbouring pixels may differ.
These small-scale differences are particularly an issue when different numbers of Gaussian components are found, and when strongly overlapping components converge to different solutions due to the large covariance in their parameters.

We account for these differences by comparing each spectrum to the properties of its nearest neighbours.
We check and attempt to correct for differences by:
\begin{enumerate}
    \item We refit the spectrum using the model with the lowest BIC from the nearest 8 pixels.
    We replace the spectrum's fit when the neighbouring model at pixel $k$ decreases the BIC significantly compared to the original model at pixel $l$, where we define ``significant'' as $\Delta {\rm BIC}_{k, l} > 10$ \citep{kassraftery_bicdiff}.
    
    \item We also refit and compare the spectrum's fit when any of the neighbouring 8 pixels has a different number of Gaussian components.
    In this case, we refit the spectrum using the models with the lowest and highest number of components in the neighbourhood, unless either is the spectrum's original model.
    We use this procedure for any difference in the number of components.
    We then re-impose the selection criteria from \S\ref{subsub:initial_fit} and replace the solution until the difference in BIC of the fits converges, i.e., $\Delta {\rm BIC}_{N_i, N_f} < 10$.
    Where applicable, both steps are applied.
    
\end{enumerate}

We apply these checks in a forward and reverse direction, looping through the spatial positions of valid fits along rows of Right Ascensions, followed by column in Declination.

We find that this procedure encourages coherent spatial solutions with a consistent number of components in the model.
This procedure is particularly important for spectra with weak signal relative to the noise, where the initial component guesses from \S\ref{subsub:component_guesser} are more susceptible to noise.
Though we do not make use of this product here (\S\ref{sub:excludecomponents}), the models after this neighbour comparison produce a far more complete and coherent model of the Milky Way \hi foreground in the M31 field.
Similar neighbour comparisons have been used in other multi-component Gaussian modeling \citep{Haud2000A&A...364...83H}, in particular, we highlight that our fitting procedure has many similarities to the \textsc{gausspyplus} algorithm \citep{RienerKainulainen2020A&A...633A..14R}.

Finally, we note that while this approach produces more spatially coherent solutions, the spectra are still fit independently of neighbouring pixels and so the spatial continuity is not explicit in the fitting.
Correctly accounting for these differences requires the model to explicitly include its nearest neighbours and be fit simultaneously \citep{Henshaw:2016el,MarchalMiville-Deschenes2019A&A...626A.101M}.
This simultaneous spatio-spectral modelling is critical when modelling blended components.
However, the large data cubes we use here would require significant computational time to model in this manner.
We will explore more coherent fitting methods in future work, where the spatial info becomes crucial for constraining blended components.

The multi-Gaussian models following the neighbour comparisons are the final models we use in our model comparisons (\S\ref{sec:mod_comp}).

\subsection{Excluding foreground and off-rotation components}
\label{sub:excludecomponents}

M33 and, primarily, M31 overlap spectrally with Milky Way \hi emission and some ``off-rotation'' emission components in both galaxies \citep[e.g., high velocity clouds;][]{Kam2017AJ....154...41K, Koch2018MNRAS}.
These features are contaminants for the model comparison we propose here since only the multi-Gaussian model can account for them.

We remove Milky Way foreground and off-rotation \hi features in two ways.
First, we assume the multi-Gaussian model correctly accounts for all detected components, regardless of their origin.
We then remove components from the multi-Gaussian model which deviate from the centroid velocity by a velocity offset $\Delta v$.
Due to the different galaxy inclinations, we choose different $\Delta v$ for our targets: $\Delta v=50$~\kms for M33 (lower inclination) and $\Delta v=80$~\kms for M31 (higher inclination).
We find that these choices appropriately remove off-rotation features in M33 \citep{Koch2018MNRAS} and remove the majority of Milky Way foreground in M31.

Second, we impose an additional masking to M31 where the red-shifted edge of the disc clearly blends with the Milky Way foreground and is not distinguished by the velocity range cut applied in the first step.
We use the interactively-selected M31 mask described in \S\ref{sec:observations} to remove components whose velocity centre is outside of the mask.
Our results in \S\ref{sec:mod_comp} do not show a systematic trend near the red-shifted edge of M31's emission suggesting this separation criterion effectively excludes foreground \hi.
The spatial footprint of our M33 observations do not include spectra that are strongly blended with Milky Way \hi, and so this second step is not applied to the M33 fits.

This separation is imposed on all results in \S\ref{sec:mod_comp}.
After this separation, we recompute the BIC statistic for the multi-Gaussian fits without the foreground/\linebreak[0]{}off-rotation components included in the model.
These recomputed BIC values are used in all of our results presented in \S\ref{sec:mod_comp}.

We do not, however, recompute the BIC values for the opaque \hi fits.
In the vast majority of the fits, the BIC statistic did not vary significantly.
The lack of change is due to the original fits not accounting for off-rotation features, since the opaque model is limited to a single component.
Because the revised multi-Gaussian models similarly do not include these features, the BIC values are similarly biased.

In the following section, we adopt the BIC \emph{difference} ($\Delta {\rm BIC}$) as the metric for comparison between the models.
The removal of foreground/\linebreak[0]{}off-rotation components in the multi-Gaussian model removes a source of bias when comparing the BIC with the opaque \hi model.
Because the opaque \hi model is limited to a single component in the fit, the model is never able to account for these off-rotation components far from the bright \hi emission.
Thus, the removal of these foreground/\linebreak[0]{}off-rotation components produces $\Delta {\rm BIC}$ a fairer comparison for the opaque \hi model.

\section{Which model is preferred for \texorpdfstring{\hibold}{HI} spectra?}
\label{sec:mod_comp}

In this section, we describe the model selection test we use to compare fits between the multi-Gaussian and opaque models, synthetic tests of the model selection test, and the results of the test.

\subsection{Model comparison using BIC}
\label{sub:model_selection}

We compare the opaque \hi and multi-Gaussian models by examining the difference in the BIC statistic (Equation~\eqref{eq:bic}; $\Delta {\rm BIC}$) of both fits to each spectrum.
Specifically, we define the BIC difference as the multi-Gaussian BIC subtracted from the opaque BIC:
\begin{equation}
    \label{eq:bic_diff}
    \Delta {\rm BIC} = {\rm BIC}_{\rm Gauss} - {\rm BIC}_{\rm Opaque}.
\end{equation}
Since a minimum BIC is optimal (Equation~\eqref{eq:bic}), $\Delta {\rm BIC} > 0$ indicates a preference for the opaque model and $\Delta {\rm BIC} < 0$ shows a preference for the multi-Gaussian model.
Following \S\ref{sub:excludecomponents}, the BIC for the multi-Gaussian models excludes foreground and off-rotation components.

Similar to selecting the number of components for the multi-Gaussian model (\S\ref{sub:gauss_model}), we choose the BIC statistic for comparing the models since it penalizes additional free parameters more strongly than other similar statistics (e.g., the Akaike Information Criteria; AIC).
Specifically, additional free parameters increase the BIC value in the $\mathrm{ln}(m)k$ term from Equation~\eqref{eq:bic}, where $m$ is the number of velocity channels and $k$ is the number of free parameters in the model.
Because each Gaussian component adds an additional 3 free parameters, the BIC is more likely to prefer a simpler model than the AIC, though the statistics will likely prefer a similar model in many cases.

Following \citet{kassraftery_bicdiff}, we consider $|\Delta {\rm BIC}| > 10$ to be strong evidence for a particular model, with smaller values indicating a weak preference that may be spurious.
Many of the comparisons exceed $|\Delta {\rm BIC}| > 10$ and so our results are not affected by the chosen threshold.

\subsection{Producing a sample of synthetic spectra}
\label{sub:synthpopulation_methods}

The model comparison we outline in the previous section is a relative comparison between two specific models.
While the multi-component Gaussian model can reproduce a wide range of spectral shapes, the single component in the opaque \hi model (Equation~\eqref{eq:thickhi_profile}) is far more limited.
A critical check for our model selection test is then to ensure that a spectrum drawn from the opaque \hi model is preferred by the model selection test.
We expect this to be true since a ``top-hat'' opaque \hi model has only 4 free parameters, while the multi-component Gaussian model would require at least 2 components, and therefore $>6$ free parameters.
However, the noise level in the data results in scatter of the BIC statistics.

To ensure our model selection test is sensitive enough to prefer true opaque \hi spectra, we produce a population of $20{,}000$ synthetic spectra using the opaque model (Equation~\eqref{eq:thickhi_profile}) with randomly drawn parameters within the parameter range used for the opaque \hi observational fits that ensure a minimum peak signal-to-noise of $>3$. 
For each random draw of parameters, we evaluate the model and fit the synthetic spectrum in the following steps:
\begin{enumerate}
    \item We produce the true model with Equation \eqref{eq:thickhi_profile}, parameterizing the optical depth with Equation \eqref{eq:tau_profile_simple}, evaluated over a spectral axis from $-200$~\kms to $+200$~\kms with $0.1$~\kms channels.
    All synthetic spectra are centred at $0$~\kms since the line centre does not affect the model comparison.
    \item We then downsample the model by averaging over $0.42$~\kms to match the observations, which produces an idealized observed spectrum \citep{KochRosolowsky2018RNAAS...2..220K}.
    \item We add Gaussian noise to the downsampled model with $\sigma_{\rm rms}=2.8$~K, matching the noise in the M31 \hi cube.
    This produces the equivalent observed spectrum to be fit.
    \item Both models are fit to the synthetic observed spectrum using the same parameters and limits used for the observational fits.
\end{enumerate}

Following the BIC model comparison from \S\ref{sub:model_selection}, we calculate the fraction of synthetic spectrum fits that strongly prefer one model ($|\Delta {\rm BIC}| > 10$).
Since the models are drawn only from the opaque model (Equation~\eqref{eq:thickhi_profile}), the model selection test will have sufficient power if the opaque model is preferred.

Table~\ref{tab:BIC_fractions} shows the break-down of model preference for the synthetic spectra.
We find that $49.0\%$ correctly prefer the opaque model ($\Delta {\rm BIC}>10$), and just $1.1\%$ of the fits strongly prefer the multi-Gaussian model ($\Delta {\rm BIC}<-10$).
The remaining $49.9\%$ of the synthetic spectrum fits fall into the ambiguous region with $-10 < \Delta {\rm BIC} < 10$ where the fits are roughly equivalent (e.g., both models converge to a single optically thin Gaussian).
A large fraction of the model comparisons remain in this ambiguous region because the opaque model requires a sufficient number of channels across the opaque or ``tophat'' shape to strongly constrain the fits.
The strongest model preference for the opaque model we find results from a combination of both high S/N \emph{and} a large line width ($\sigma>15$~\kms).
For spectra with these wide line widths, the multi-Gaussian model typically infers $>1$ components.
If we consider only the synthetic spectra fit with $>1$ components in the multi-Gaussian model, we find that $80.4\%$ of the spectra prefer the opaque model ($\Delta {\rm BIC}>10$), $18.0\%$ are ambiguous ($-10 < \Delta {\rm BIC} < 10$) and $1.5\%$ prefer the multi-Gaussian model ($\Delta {\rm BIC}<-10$).
This demonstrates that the preference becomes much stronger for the opaque model fit when multiple Gaussian components are inferred.


These results confirm that our fitting approach and model selection test has sufficient power to distinguish between the opaque and multi-Gaussian models in our \hi observations, and that that power diminishes as the signal-to-noise in the spectrum decreases, as expected.
In the following sections, we use these results from the synthetic spectra to compare and contrast with the model preference from the observed \hi spectra.

\subsection{A strong preference for multi-Gaussian models}
\label{sub:model_tests}

Using the comparison framework from \S\ref{sub:model_selection}, we examine $\Delta {\rm BIC}$ for the population of \hi spectra in M31 and M33.
Our comparisons in this section demonstrate a strong preference for a multi-Gaussian model.

\begin{figure*}
\includegraphics[width=\textwidth]{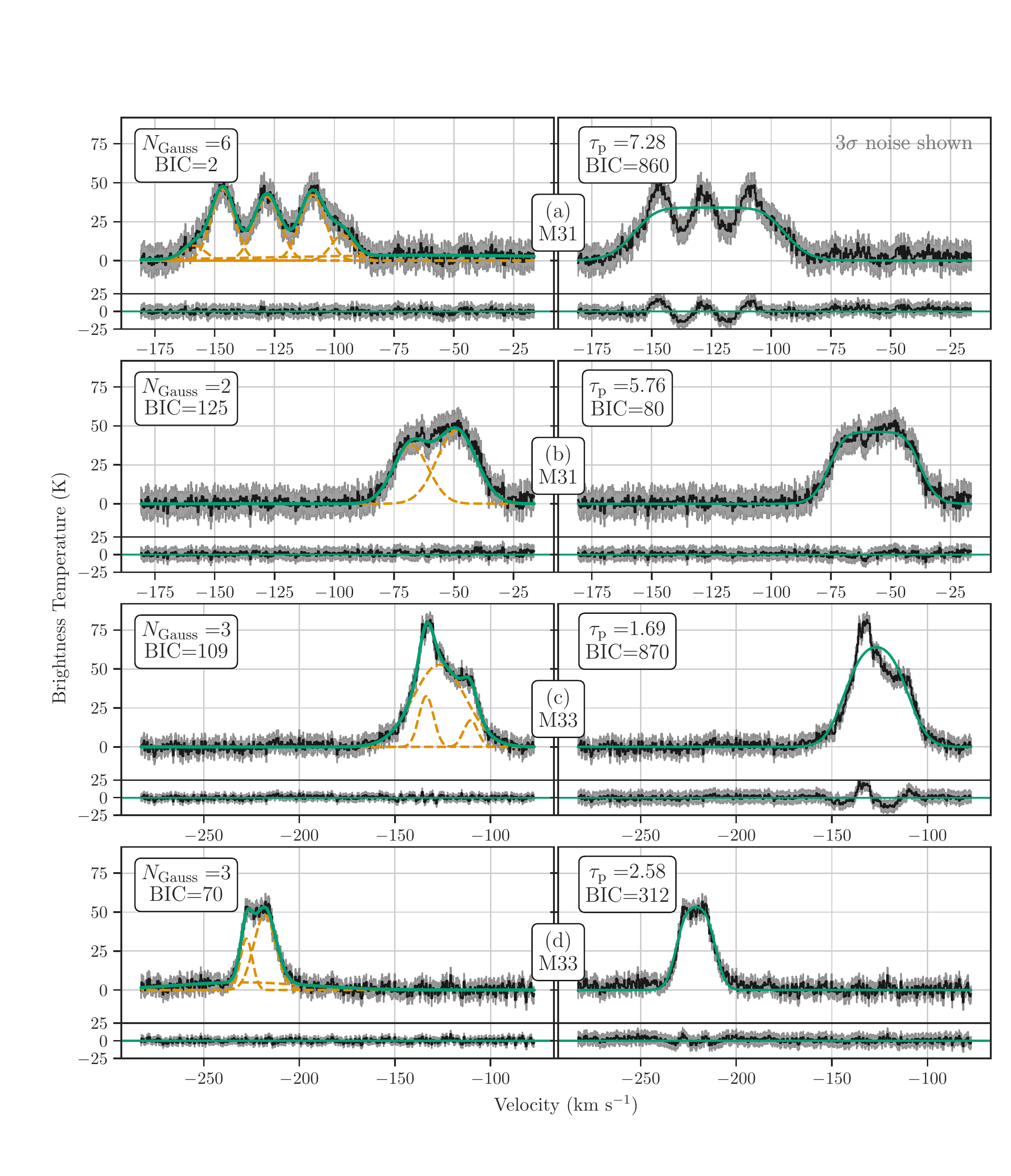}
\vspace*{-35px}
\caption{\label{fig:m31_m33_spectra_wfits}
Four of the \hi spectra shown in Figure~\ref{fig:m31_m33_spectra} with their multi-Gaussian (left) and opaque (right) model fits.
The fit residuals are shown in the panel below each spectrum and fit, with the $3\sigma$ rms noise shown in gray.
The number of Gaussians and the peak opacity ($\tau_{\rm p}$) are stated for each model, respectively, as well as the BIC statistic for each fit.
In three of the examples (a, c, and d), the BIC statistic is smaller for the multi-Gaussian model and is therefore the preferred model.
The opaque model infers $\tau_{\rm p}>1$ in all three cases, but the fit quality is noticeably worse, especially for (a) and (c) where there are clearly multiple components.
Panel (b) shows a case where the opaque and multi-Gaussian models provide a similar fit to the data (i.e., the fit residuals are similar), but the opaque model has fewer free parameters and the smaller BIC values indicates it is preferred.
We assess the validity of this visual inspection by comparing the BIC statistics for the entire population of fitted spectra.
}
\end{figure*}

We first visually examine the fits for a few spectra, chosen specifically to highlight a range in the inferred peak optical depth from the opaque model.
Figure~\ref{fig:m31_m33_spectra_wfits} shows four example spectra with fits to both models (chosen from a subset of the spectra in Figure~\ref{fig:m31_m33_spectra}).
Each panel in the figure shows the key fit results, including the number of Gaussians ($N_{\rm Gauss}$), inferred peak optical depth ($\tau_{\rm p}$), and the BIC values are stated for each fit.
Following visual inspection, the spectra in panels (a) and (c) clearly demonstrate multiple peaks and strongly prefer the multi-Gaussian model based on their smaller BIC values ($\Delta {\rm BIC} < 0$).
The spectra in panels (b) and (d) both have a single bright ``pedestal,'' which is qualitatively the flat-top shape expected for the opaque \hi model over the velocity range where $\tau>1$ \citep{Braun2009ApJ...695..937B}
Only spectrum (b) prefers the opaque model ($\Delta {\rm BIC} > 0$), however.
In panel (d) the multi-Gaussian model fit has a much smaller BIC value and is therefore preferred..

\begin{figure*}
\includegraphics[width=\textwidth]{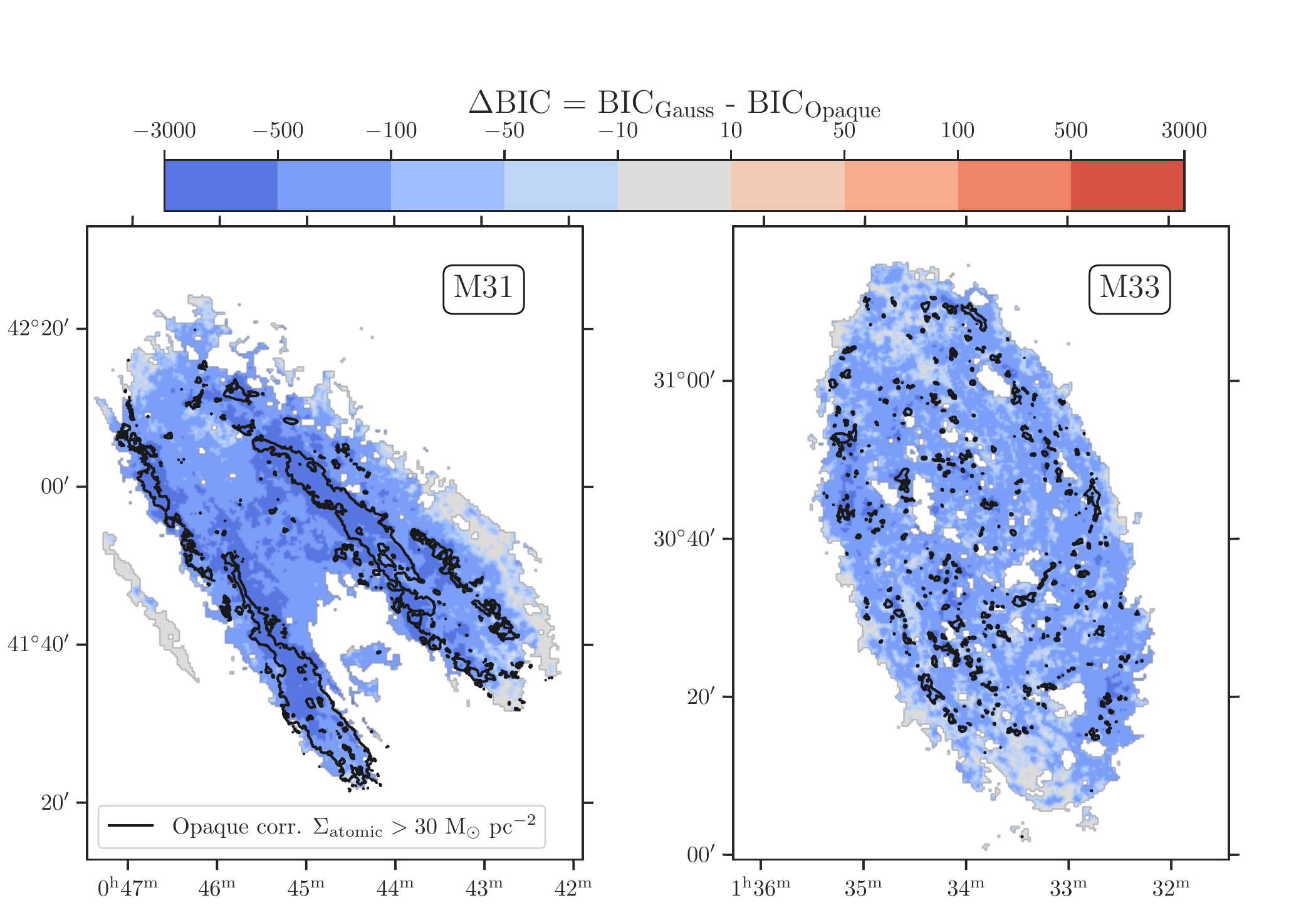}
\caption{\label{fig:m31_m33_delta_bic_map}
$\Delta$BIC values comparing the multi-Gaussian and opaque models.
Negative values (blue) indicate a preference for the multi-Gaussian model, while positive values (red) show where the optically thick model is preferred.
Gray regions in the map indicate where neither model has a strong model preference ($|\Delta {\rm BIC}| < 10$).
The black contours show where $\Sigma_{\rm atomic} > 30$~\msolpcsq \textit{with the opacity correction from Equation~\eqref{eq:thickhi_profile}}.
This surface density exceeds all values assuming optically thin emission.
Multi-Gaussian models are preferred in the majority of spectra.
Further, as the contour regions indicate, regions with the strongest preference for multi-Gaussian (largest negative $\Delta$BIC) correlate with the largest inferred opacity corrections from the opaque model.
}
\end{figure*}

To test this apparent preference for the multi-Gaussian model, we create $\Delta {\rm BIC}$ maps of both galaxies.
Figure~\ref{fig:m31_m33_delta_bic_map} shows $\Delta {\rm BIC}$ in M31 and M33 where valid fits are found (\S\ref{sec:hi_models}).
The colour is centered such that $\Delta {\rm BIC}=0$ is shown in gray.
Throughout the maps, we find a strong preference for the multi-Gaussian model ($\Delta {\rm BIC} < 0$).
Most locations where the opaque model is preferred ($\Delta {\rm BIC} > 0$) correspond to regions with faint \hi emission (Figure~\ref{fig:m31_m33_maps}).
This result is in contrast with the findings by  \citet{Braun2009ApJ...695..937B} and \citet{Braun2012ApJ...749...87B}, where opaque \hi is found in regions of bright \hi  \citep[e.g.,][]{Braun1997ApJ...484..637B}.
While we find that the largest opacity corrections correspond to regions with bright \hi, as shown by the black contour in Figure~\ref{fig:m31_m33_delta_bic_map}, these are also the spectra that tend to have the strongest preference for the multi-Gaussian model.

We see a general trend towards $\Delta {\rm BIC} \approx 0$ near the edge of \hi detections in both maps, where the noise increases (e.g., in M31) and the \hi emission is fainter.
At lower signal-to-noise, the multi-Gaussian model tends to have a single component, and therefore is similar to the opaque model in the optically thin limit.
In these regions, there is a small tendency for $\Delta {\rm BIC}<0$ since a single Gaussian component has three free parameters while the optically thin limit of the opaque model includes a fourth unconstrained parameter, \tspin, and so the former model is preferred.

\begin{table*}
    \centering

\begin{tabular}{cllll}
    & &  Multi-Gaussian preferred & No strong preference & Opaque preferred \\\
    & &  \% with $\Delta {\rm BIC} < -10$ & \% with $-10 < \Delta {\rm BIC} < 10$ & \% with $\Delta {\rm BIC} > 10$ \\\hline
M31    & All (Figure \ref{fig:delta_bic_vs_darknhi}; top) &  $82.0$ & $17.5$ & $0.5$ \\
       & $\tau_{\rm p} > 1$ (Figure \ref{fig:delta_bic_vs_darknhi}; bottom) & $88.7$  & $10.5$ & $0.8$ \\
       & $\tau(v) > 0.5$ (Figure \ref{fig:delta_bic_vs_darknhi_taulimonly}) & $83.7$ & $15.4$ & $0.9$ \\[6px]
M33    & All (Figure \ref{fig:delta_bic_vs_darknhi}; top) &  $80.2$ & $19.6$ & $0.2$ \\
       & $\tau_{\rm p} > 1$ (Figure \ref{fig:delta_bic_vs_darknhi}; bottom) & $83.2$ & $16.0$ & $0.7$ \\
       & $\tau(v) > 0.5$ (Figure \ref{fig:delta_bic_vs_darknhi_taulimonly}) & $74.4$ & $25.0$ & $0.6$ \\[6px]
Synth.    & All (Figure \ref{fig:delta_bic_vs_darknhi}; top) &  $1.1$ & $49.9$ & $49.0$ \\
          & $\tau_{\rm p} > 1$ (Figure \ref{fig:delta_bic_vs_darknhi}; bottom) & $1.4$ & $35.0$ & $63.6$ \\
          & $\tau(v) > 0.5$ (Figure \ref{fig:delta_bic_vs_darknhi_taulimonly}) & $0.6$ & $78.5$ & $20.9$ \\
\end{tabular}

    \caption{Percentage of spectra for different ranges in $\Delta {\rm BIC}$ in the M31, M33, and synthetic spectra fit distributions.
    We define three ranges: (i) a strong preference for the multi-Gaussian model ($\Delta {\rm BIC} < -10$); (ii) spectra without a strong preference for either model ($-10 < \Delta {\rm BIC} < 10$); and (iii) a strong preference for the opaque model ($\Delta {\rm BIC} > 10$).
    We compute the percentage in the ranges for the three tests including all fits and only those with $\tau_{\rm p} > 1$ (Figure~\ref{fig:delta_bic_vs_darknhi}), and the BIC computed only where $\tau(v) > 0.5$ (Figure~\ref{fig:delta_bic_vs_darknhi_taulimonly}).
    Our synthetic spectrum fit distribution correctly shows a preference for the opaque model, from which the spectra are drawn from.
    However, the vast majority of observational fits in M31 and M33 prefer the multi-Gaussian model for all three comparisons.}
    \label{tab:BIC_fractions}
\end{table*}

Table~\ref{tab:BIC_fractions} shows the fraction of spectra prefer each model, split into categories of $\Delta {\rm BIC} < -10$, $-10<\Delta {\rm BIC} < 10$, and $\Delta {\rm BIC} > 10$, as defined in \S\ref{sub:model_selection}.
As shown in the maps of $\Delta {\rm BIC}$ in Figure~\ref{fig:m31_m33_delta_bic_map}, we find that most ($82.0\%$ and $80.2\%$) of the spectra prefer the multi-Gaussian model, while just $0.5\%$ and $0.2\%$ prefer the opaque model in M31 and M33, respectively.
These fractions differ significantly from the synthetic sample of opaque fits, where $49.0\%$ prefer the opaque model.

To visualize which spectra most strongly prefer the multi-Gaussian model, we compare the model preference $\Delta {\rm BIC}$ to how much the opaque \hi model deviates from a single Gaussian (i.e., the optically thin limit).
This deviation of the opaque \hi model from a Gaussian is difficult to parameterize in terms of a single fit parameter, or, for example, the peak optical depth.
The reason for this is because the three parameters that determine the profile shape---$T_{\rm p}$, $T_{\rm s}$, and $\sigma$---are highly correlated.
Instead, we compare $\Delta {\rm BIC}$ to the the apparent ``dark'' \hi intensity, $I_{\rm dark}$, which is the integrated difference of the opaque model fit (Equation \eqref{eq:thickhi_profile}) if the medium were optically thin and the model itself:
\begin{equation}
    \label{eq:darkhi}
    I_{\rm dark} = \int \left[ T_{\rm p} \textrm{exp}\left(- \frac{(v - v_{\rm 0})^2}{2\sigma_{\rm i}^2} \right) - T_{\rm s} \left( 1 - {\rm exp}\left[-\tau(v) \right] \right) \right] {\rm d}v,
\end{equation}
where $\tau(v)$ is given by Equation \eqref{eq:tau_profile_simple}.
When the model fit infers the spectrum is optically thin, $I_{\rm dark}$ converges to $0$.
We calculate the $1\sigma$ uncertainty on $I_{\rm dark}$ using the \textsc{uncertainties} package\footnote{v3.1.5; \url{https://pythonhosted.org/uncertainties/}} which propagates the parameter uncertainties from the opaque model fit (Equation \ref{fig:opaque_model}) in Equation \ref{eq:darkhi}.
We note that these uncertainties, following the fit parameter uncertainties, assumes they are normally-distributed.
This assumption will break-down when $I_{\rm dark}$ approaches the optically thin limit because $I_{\rm dark}$ cannot be negative.

We stress that interpreting $I_{\rm dark}$ as a physical quantity depends on the opaque model fitting a spectrum well.
As we show above, $<1\%$ of the observed spectra prefer the opaque model, and so the opaque fit parameters should be considered carefully before assigning a physical meaning to what may be a poor fit.
Because of this, we consider $I_{\rm dark}$ as a useful \emph{metric} to compare with, rather than a physical quantity.

\begin{figure*}
\includegraphics[width=\textwidth]{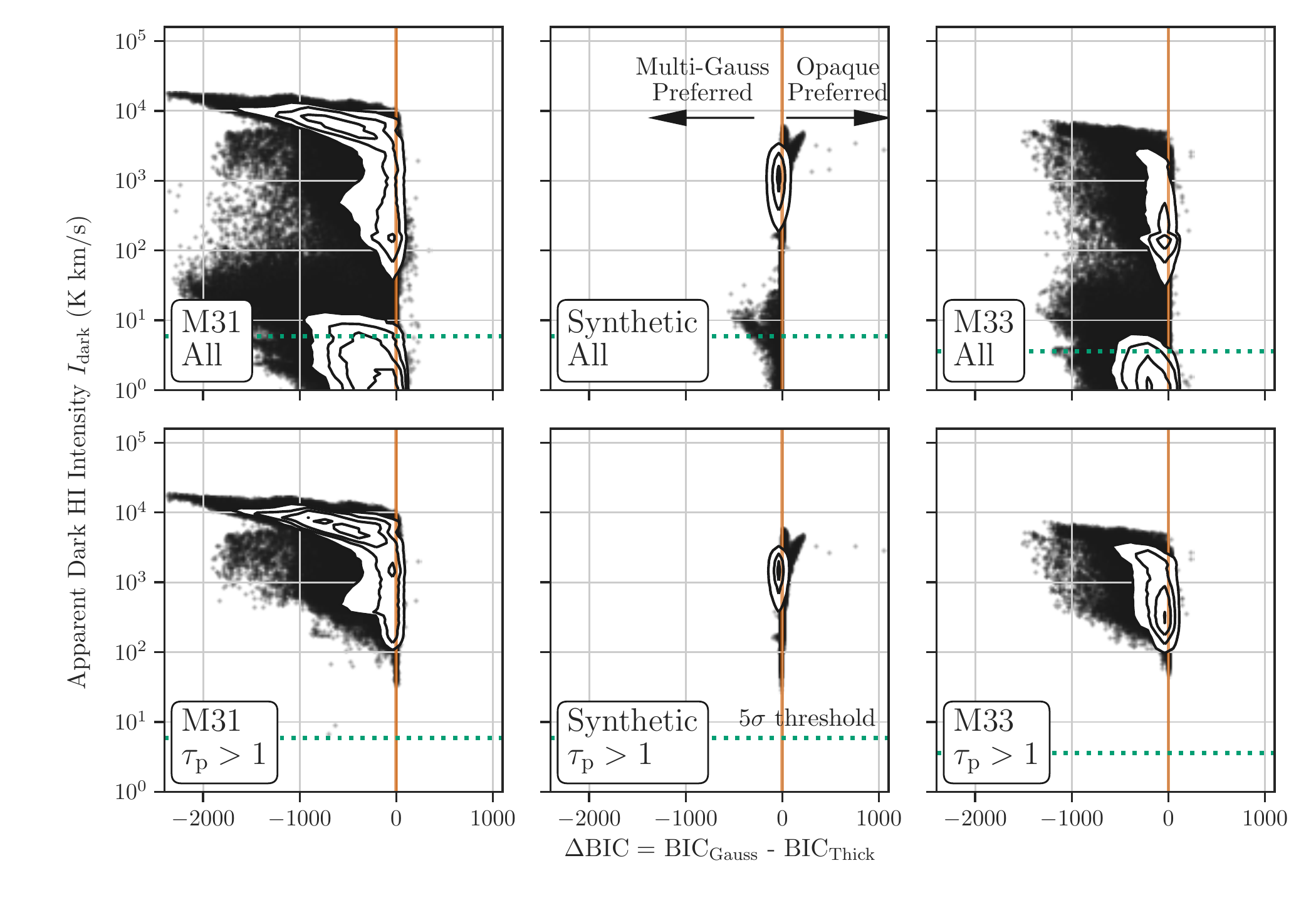}
\vspace*{-20px}
\caption{\label{fig:delta_bic_vs_darknhi}
Apparent dark \hi integrated intensity from the opaque model (Equation~\eqref{eq:darkhi}; i.e., optically thin equivalent line subtracted by the optical depth-corrected profile) plotted against $\Delta{\rm BIC}$.
$\Delta{\rm BIC} < 0$ indicates a preference for the multi-Gaussian model, while $\Delta{\rm BIC} > 0$ are fits with the opaque model preferred.
The panels show the plots for M31 (left), synthetic spectra generated from Equation \eqref{eq:thickhi_profile} (centre), and M33 (right).
The top row shows the results for all spectra, while the second row shows fits where the peak optical depth from the opaque model fit is $\tau_{\rm p} > 1$.
The contours in each panel show the density of fits contained at the $1{-}4\sigma_{\rm rms}$ levels, while individual data points show individual fits that fall outside of the $4\sigma_{\rm rms}$ contour.
The narrow vertical shaded region indicates where $-10 < \Delta {\rm BIC} < 10$ where there is no strong preference for either model.
The horizontal dotted line is the $5\sigma_{\rm rms}$ sensitivity in integrated intensity for a single $0.42$~\kms spectral channel.
The observed fit results strongly prefer a multi-Gaussian model ($\Delta {\rm BIC}<0$), while the synthetic sample correctly prefers the opaque model used to generate them ($\Delta {\rm BIC}>0$).
These differences become even more apparent when \emph{only} opaque model fits are shown in the second row, demonstrating the opposite trend expected for fits to the opaque model.
Despite having many more free parameters, the multi-Gaussian model is consistently preferred.
}
\end{figure*}

Figure~\ref{fig:delta_bic_vs_darknhi} shows the $\Delta {\rm BIC}$ statistic distributions for M31, M33, and the synthetic spectrum fits plotted against $I_{\rm dark}$. 
The top row shows the distribution of all fitted spectra.
Consistent with the maps in Figure~\ref{fig:m31_m33_delta_bic_map}, the \hi spectra in M31 and M33 with the strongest preference for the multi-Gaussian tend to have a larger $I_{\rm dark}$ inferred from the opaque model.
This trend is the opposite of the trend for our fits to the synthetic spectra of the opaque model (\S\ref{sub:synthpopulation_methods}; middle panel in Figure~\ref{fig:delta_bic_vs_darknhi}), where the preference for the opaque model in several fits \emph{increases} at large $I_{\rm dark}$.

A common trend of all three samples (M31, M33, and the synthetic spectra) in Figure~\ref{fig:delta_bic_vs_darknhi} is the broad distribution of $\Delta {\rm BIC}$ values when $I_{\rm dark}$ approaches zero.
This broad range shows where the opaque model approaches the optically thin solution, where some spectra are better modelled with multiple components or where the signal-to-noise is low and both models approach a similar fit, as we see at the edges of the maps in Figure~\ref{fig:m31_m33_delta_bic_map}.

We next examine whether the preference for a multi-Gaussian model persists if we consider only the spectra where opaque \hi emission is inferred from the opaque model, or equivalently where the peak opacity is $\tau_{\rm p} > 1$.
After applying this criterion, we find that the same general trend persists: most spectra prefer the multi-Gaussian model and that preference becomes stronger for the largest inferred $I_{\rm dark}$.
We find that fraction of spectra that prefer the multi-Gaussian model increase to $88.7\%$ for M31 and $83.2\%$ for M33.
This comparison is shown in the bottom row of Figure~\ref{fig:delta_bic_vs_darknhi}.
Because only a sub-set of the $\Delta {\rm BIC}$ are shown, the percentiles defining the contours in the plots are changed, leading to some difference in the observed distributions.

This trend of a stronger preference for the multi-Gaussian model where opaque \hi is inferred from the opaque model (Equation~\eqref{eq:thickhi_profile}) is opposite of the trend with the synthetic spectra.
In the synthetic sample, we find that the fraction of spectra that prefer the opaque model increases from $49.0\%$ to $63.6\%$ when considering only spectra where opaque \hi is inferred from the opaque model (Table~\ref{tab:BIC_fractions}).
This follows our expectation for the opaque model since the spin temperature \tspin is only constrained when the line shape approaches a ``top-hat'' shape (Figure~\ref{fig:opaque_model}).

Finally, we include one further test that explicitly accounts for the single-component nature of the opaque model (Equation~\eqref{eq:thickhi_profile}).
The spectrum in panel (d) in Figure~\ref{fig:m31_m33_spectra_wfits} demonstrates the limits of the opaque model well.
The multi-Gaussian model correctly includes a wide component that accounts for faint ``tails'' in the emission.
However, the opaque model cannot account such features and is therefore more likely to have a larger BIC, even if the model has fewer parameters.

To address this limitation, we recalculate the BIC statistic for both models limited to the channels where $\tau(v) > 0.5$.
In terms of the opaque model, this limits the model selection to comparing \emph{only} the spectral range inferred to have opaque \hi emission (see the example opaque model in Figure~\ref{fig:opaque_model}) or where the profile will have a ``top-hat'' shape.
This limit is chosen based on our synthetic spectrum fits where the opaque model measurably deviates from a Gaussian line shape given the noise level and spectral resolution in our observations.
For the multi-Gaussian model, we consider only the components where $>75\%$ of their intensity is within range of $\tau(v) > 0.5$.

\begin{figure*}
\includegraphics[width=\textwidth]{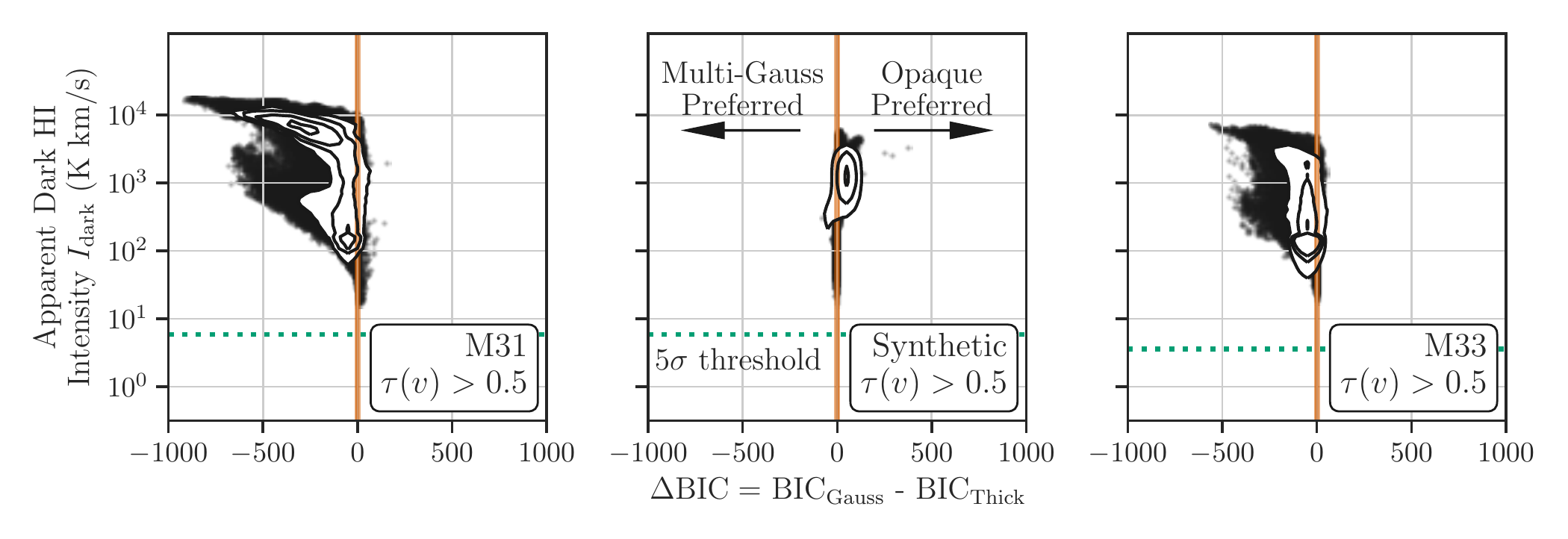}
\vspace*{-15px}
\caption{\label{fig:delta_bic_vs_darknhi_taulimonly}
$\Delta {\rm BIC}$ plotted against the apparent dark \hi integrated intensity from the opaque model, similar to Figure~\ref{fig:delta_bic_vs_darknhi}.
Here, we show only the fit statistics calculated where $\tau(v) > 0.5$ from Equation \eqref{eq:tau_profile_simple}.
This comparison controls for the single-component limitations of the opaque model, which cannot account for multiple peaks in a spectrum.
By computing the BIC statistics only where $\tau(v) > 0.5$, this comparison solely compares the models where the ``flat-top'' is measurable in the spectra.
However, these results are similar to the $\Delta {\rm BIC}$ over the whole spectra (Figure~\ref{fig:delta_bic_vs_darknhi}).
The synthetic spectra, drawn from Equation~\eqref{eq:thickhi_profile} with observational noise added, confirm that our observations are sensitive to opaque line profiles and the fitting procedures correctly find $\Delta {\rm BIC} > 0$ on average.
The observed spectra in M31 (left) and M33 (right), however, have $\Delta {\rm BIC} < 0$, indicating a continued strong preference for the multi-Gaussian model.
}
\end{figure*}

Figure~\ref{fig:delta_bic_vs_darknhi_taulimonly} shows the resulting distribution of $\Delta {\rm BIC}$ limited to $\tau(v) > 0.5$ and the model preference fractions are given in Table~\ref{tab:BIC_fractions}.
As expected, the synthetic spectrum fit distribution shows a preference for the opaque model, with $20.9\%$ spectra with $\Delta {\rm BIC} > 10$ and just $0.6\%$ with $\Delta {\rm BIC}<-10$ preferring the multi-Gaussian model.
These fractions are nearly the opposite of what we find for the observed spectra, with $0.9\%$ and $0.6\%$ of spectra with $\Delta {\rm BIC} > 10$, while $83.7\%$ and $74.4\%$ have $\Delta {\rm BIC} < -10$ for M31 and M33, respectively.

In all three distributions, a larger fraction of spectra do not show a strong preference for one model or another ($-10 < \Delta {\rm BIC} < 10$), particularly for the synthetic spectrum fit distribution which has $78.5\%$ in this ambiguous region.
This trend occurs because this test limits the number of spectral channels that are used for the model comparison, in effect, limiting the amount of information available and increasing the scatter in $\Delta {\rm BIC}$ values.
Despite the fact that the $\tau(v) > 0.5$ restriction makes $\Delta {\rm BIC}$ a noisier metric, we still find a strong preference for multi-Gaussian models in both M31 and M33.
This result demonstrates that the preference for multi-Gaussian models does not result from the opaque model being limited to a single component (Equation~\eqref{eq:thickhi_profile}).

In summary, we identify a strong preference for the multi-Gaussian model in observed \hi spectra when compared to a single opaque \hi component model.
This result indicates that most \hi are best described as emission from multiple atomic ISM components along the line-of-sight.

\subsection{Effect of spectral resolution on model preference}
\label{sub:specres_effect_on_modelpref}

\begin{figure}
\includegraphics[width=0.45\textwidth]{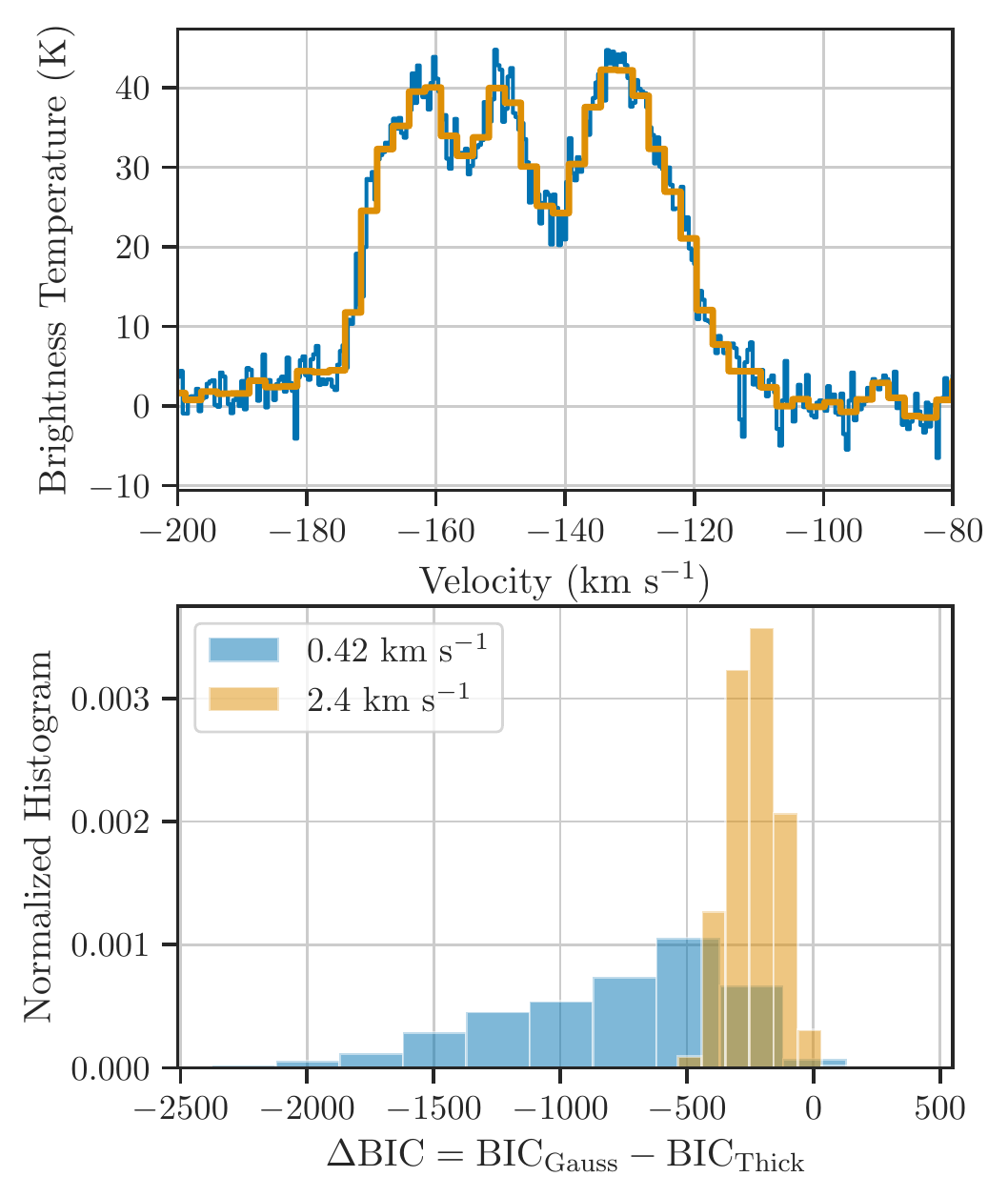}
\vspace*{-10px}
\caption{\label{fig:spectral_deltaBIC_comparison}
Effect of spectral resolution shown with an example \hi spectral shape (top) and distributions of $\Delta {\rm BIC}$ of a sub-region of M31 (bottom).
We demonstrate here that a coarser spectral resolution removes spectral shape information from the original spectra, making some components appear more blended.
In this case, we plot the spectrum with $2.4$~\kms spectral channels, similar to the resolution of the M31 \hi data from \citet{Braun2009ApJ...695..937B}.
The bottom panel shows the model selection comparison with $\Delta {\rm BIC}$ of a $200\times200$~pixel region in our M31 \hi data.
Though the multi-Gaussian model is preferred in both cases, we find that fits to the coarser resolution data increase most $\Delta {\rm BIC}$ values, indicating that the model preference is \emph{weaker} for spectra with coarser spectral resolution.
}
\end{figure}

A key difference in the analysis we perform here relative to that of \citet{Braun2009ApJ...695..937B} is the improved spectral resolution of the \hi observations.
Here, we test how changing the spectral resolution affects the model preference by selecting a $1\farcm1\times1\farcm1$ ($200\times200$~pixel) region in M31 centred at $0^{\rm h}44^{\rm m}34^{\rm s} \ {+}41^{\rm d}55^{\rm m}13^{\rm s}$.
This small region spans across where both large and small opaque regions are inferred from the opaque model fit (see Figure~\ref{fig:m31_m33_delta_bic_map}).
We spectrally smooth and then downsample the \hi spectra to have spectral channels of $2.4$~\kms, similar to the $2.3$~\kms spectral channels in the M31 \hi observations from \citet{Braun2009ApJ...695..937B}.
The top panel in Figure~\ref{fig:spectral_deltaBIC_comparison} shows an example spectrum at the original and coarser spectral resolution.

We then applied the same fitting approach to the spectrally-downsampled \hi spectra, as described in \S\ref{sec:hi_models}, and calculate the $\Delta {\rm BIC}$ between the two model fits.
In the bottom panel of Figure~\ref{fig:spectral_deltaBIC_comparison}, we plot histograms of $\Delta {\rm BIC}$ for fits at the original $0.42$~\kms and coarser $2.4$~\kms spectral resolutions.
We find that, while both distributions indicate a preference for the multi-Gaussian model ($\Delta {\rm BIC} < -10$), the preference is weaker for fits at coarser spectral resolution.
This result demonstrates that the model selection test becomes weaker as information is lost by downsampling the spectra.
For example, the example spectrum in Figure~\ref{fig:spectral_deltaBIC_comparison} has two blended peaks that become more difficult to distinguish at coarser spectral resolution.

We additionally highlight that, because we spectrally smooth the data when downsampling, the signal-to-noise is also improved relative to the original data.
If the noise levels were higher---comparable to the \hi data used by \citet{Braun2009ApJ...695..937B} and \citet{Braun2012ApJ...749...87B} (Table~\ref{tab:observe_comp})---the $\Delta {\rm BIC}$ would further increase and trend towards $\Delta {\rm BIC} \sim 0$.
Including additional noise will lead to poorer discrimination between the models.

The test we present here demonstrates a key advantage for finer spectral resolution \hi observations when the signal-to-ratio is sufficient to constrain the spectral shape.
When the spectral resolution is too coarse, leading to highly blended \hi components, it becomes more difficult to identify a model preference.
This example highlights how the trade-off between using coarser spectral resolution, to maximize signal-to-noise, versus finer spectral resolution, to best recover the spectral shape, affects our ability to model the \hi spectra.

\section{Comparison with dust tracers}
\label{sec:dustcomparisons}

Dust emission and dust extinction provide independent measurements of ISM column density that can be compared with the inferred \hi opaque corrections from the opaque model (Equation \ref{eq:thickhi_profile}).
In this section, we compare the atomic gas surface density and inferred opacity corrections to the dust surface density from modeling the IR SED in M31 and M33 from \citet{UtomoChiang2019ApJ...874..141U}, and the mean visual extinction $A_{\rm V}$ in M31 from \citet{DalcantonFouesneau2015ApJ...814....3D} over the region observed by PHAT.
We additionally correct the surface densities for disc inclination in both galaxies (see \S\ref{sub:vla}).
Finally, to emphasize the atomic gas to dust comparison, we exclude regions where the molecular gas dominates the neutral ISM.
Specifically, we remove regions with $\Sigma_{\rm mol}>1.5$~\msolpcsq based on CO emission using \coonezero in M31 \citep{NietenNeininger2006A&A...453..459N} and \cotwoone in M33 \citep{Gratier2012A&A...542A.108G,Druard2014A&A...567A.118D}, where we correct for the galaxy inclination angle and use a conversion factor of $\alpha_{\rm CO10}=4.3$~\msolconvert with a line ratio of $R_{21}=0.7$ for the \cotwoone line.
The surface density needed for \htwo formation is $\sim10$~\msolpcsq, and increases with decreasing metallicity \citep[e.g.,][]{Krumholz2009ApJ...699..850K,SchrubaBialy2018ApJ...862..110S}, so our chosen threshold should keep only regions where the atomic gas is the primary neutral ISM component.

By considering only regions where the atomic gas dominates the neutral ISM, we expect the relation between dust and atomic gas surface density to approximately be $\Sigma_{\rm dust} \approx {\rm GDR}^{-1} \Sigma_{\rm atomic}$, where ${\rm GDR}$ is the gas-to-dust ratio.
If a significant part of the \hi along a line-of-sight is opaque, following Equation~\eqref{eq:thickhi_profile}, we expect to find two features in the relation between $\Sigma_{\rm atomic}$ and $\Sigma_{\rm dust}$: (i) additional scatter when using the optically thin assumption for \hi, with a population of lines-of-sight with a low apparent GDR where $\Sigma_{\rm atomic}$ is underestimated\footnote{Despite removing lines-of-sight where molecular gas dominates, there remain many spectra with a large inferred opacity based on the opaque model fit (Equation \ref{eq:thickhi_profile}).
Thus, we would still expect to find low apparent GDR lines-of-sight if the inferred \hi opacity is accurate.}; and (ii) reduced scatter after applying the inferred \hi opacity correction as this correction accounts for a missing component of $\Sigma_{\rm atomic}$ that is traced by the dust.

\begin{figure*}
\includegraphics[width=0.8\textwidth]{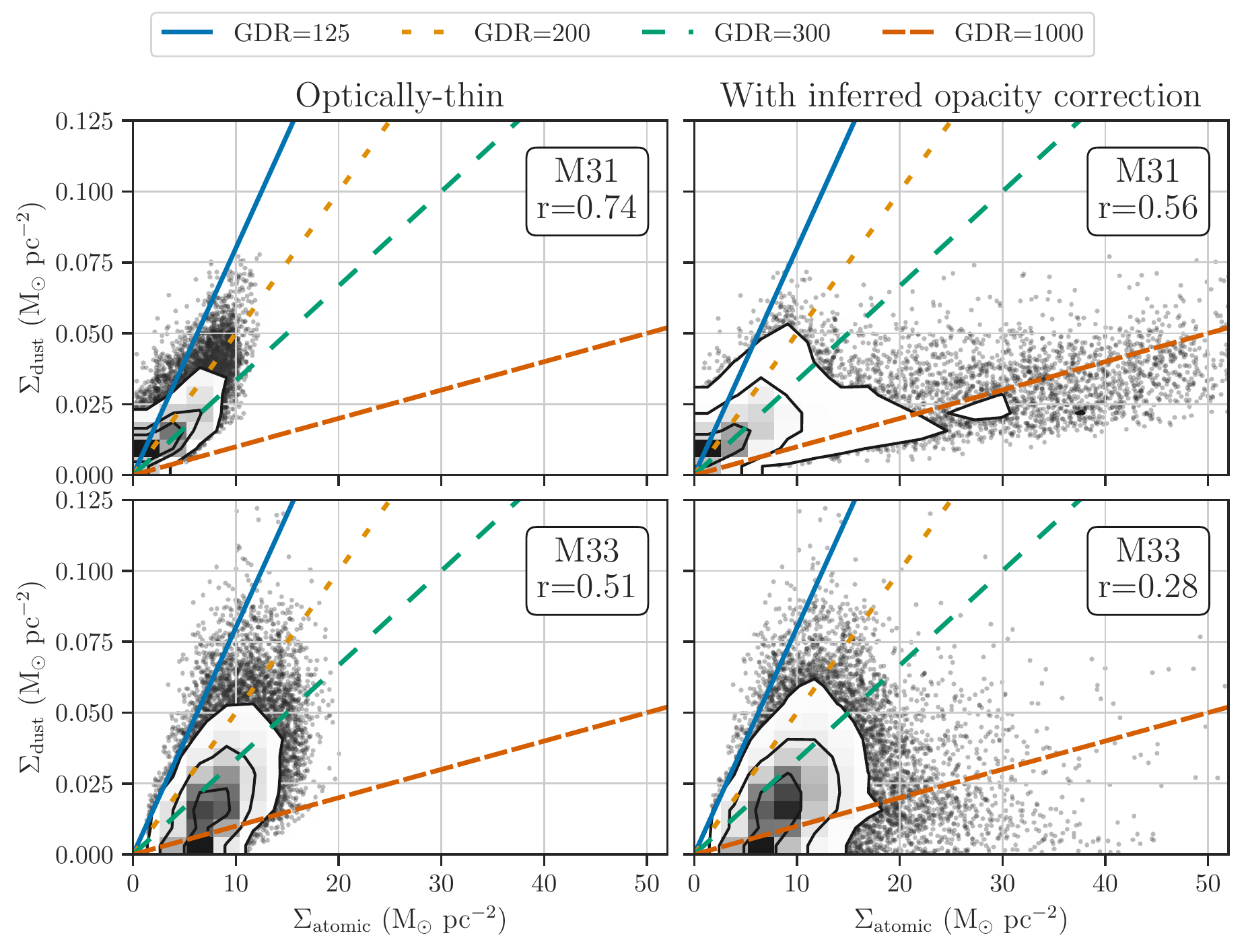}
\caption{\label{fig:dust_hi_gdr}
Atomic gas surface density versus dust surface density without (left column) and with (right column) \hi opacity corrections from Equation \eqref{eq:thickhi_profile}.
All values are corrected for the disc inclination along the line-of-sight and are convolved to a common physical resolution of 167~pc, matching the dust surface density maps from \citet{UtomoChiang2019ApJ...874..141U}.
In each panel, regions with $\Sigma_{\rm mol}>1.5$~\msolpcsq (i.e., non-negligible molecular gas mass) are removed, so that to first order the atomic and dust surface densities are related only by the gas-to-dust ratio \citep[GDR; e.g.,][]{SandstromLeroy2013ApJ...777....5S}, and so we report the linear correlation coefficient $r$ in each panel.
A range of GDRs are shown with overlaid lines.
The inferred opaque \hi corrections from the opaque model (Equation \ref{eq:thickhi_profile}) significantly increase the scatter in this relation for both galaxies, and further require a factor of $\sim10$ range in the ${\rm GDR}$ to explain.
Consistent with our spectral model comparisons, these results suggest that the opaque model overestimates or has highly uncertain opaque \hi corrections.
}
\end{figure*}

Figure~\ref{fig:dust_hi_gdr} shows the atomic gas versus dust surface density in M31 and M33, with and without the \hi opacity correction.
We smooth the \hi surface density maps to match the $167$~pc resolution of the dust surface density from \citet{UtomoChiang2019ApJ...874..141U}, which corresponds to $46\farcs3$ for M31 and $41\arcsec$ and for M33.
Using the optically thin \hi assumption, we find a strong correlation between $\Sigma_{\rm atomic}$ and $\Sigma_{\rm dust}$ consistent with ${\rm GDR}\approx100{-}300$.
This range in GDR broadly agrees with previous findings in M31 and M33 \citep[e.g.,][]{NietenNeininger2006A&A...453..459N,LeroyBolatto2011ApJ...737...12L,GratierBraine2017A&A...600A..27G,WilliamsGear2019MNRAS.483.5135W}.
We note that most studies also include molecular gas as traced by CO and so do not remove molecular gas dominated lines-of-sight.

Including the inferred \hi opacity corrections from the opaque model (Equation \ref{eq:thickhi_profile}), the correlation between $\Sigma_{\rm dust}{-}\Sigma_{\rm atomic}$ is reduced and the scatter in the relation significantly increases.
In both galaxies, lines-of-sight with large \hi opacity corrections require a much larger ${\rm GDR}$ to explain the observed relations.
Particularly in M31, regions with large \hi opacity corrections require ${\rm GDR} \approx 1000$.
Higher values of the ${\rm GDR}$ are expected in low-metallicity galaxies, such as the SMC \citep{LeroyBolatto2011ApJ...737...12L,Roman-DuvalGordon2014ApJ...797...86R}, but are otherwise not found in higher metallicity galaxies in the Local Group \citep{LeroyBolatto2011ApJ...737...12L} or other nearby galaxies \citep{SandstromLeroy2013ApJ...777....5S}.

We can further compare the \hi surface density with the dust extinction $A_{\rm V}$ in M31 \citep{DalcantonFouesneau2015ApJ...814....3D}, which provides another useful measure of the dust with independent systematics from the IR SED fitting \citep[e.g.,][]{WhitworthMarsh2019MNRAS.489.5436W}.
The dust extinction provides two key advantages relative to the dust surface density from fitting the dust emission SED.
First, $A_{\rm V}$ is not dependent on the dust temperature.
This overcomes ambiguity when comparing with the dust emission SED, where the dust surface density and temperature are correlated.
This correlation could explain the large range in DGR since the SED fit is a mixture of warm and cold dust along the line-of-sight and the warm component, which radiates more strongly per unit mass, biases the inferred dust temperature \citep{SandstromLeroy2013ApJ...777....5S}.
Second, the IR bands, where the dust SED is constrained, have a coarser spatial resolution compared to our \hi data by a factor of $\sim2$.
The M31 $A_{\rm V}$ map can be compared at the same spatial resolution as the \hi maps and thereby tests whether compact high surface density regions are smoothed out at the coarser resolution of the dust emission maps.
We note, however, that we lack matched spatial resolution CO data covering the entire \hi mosaic (Figure~\ref{fig:m31_m33_maps}).
Because of this, we do \textit{not} exclude molecular dominated regions in this comparison.

\begin{figure*}
\includegraphics[width=0.8\textwidth]{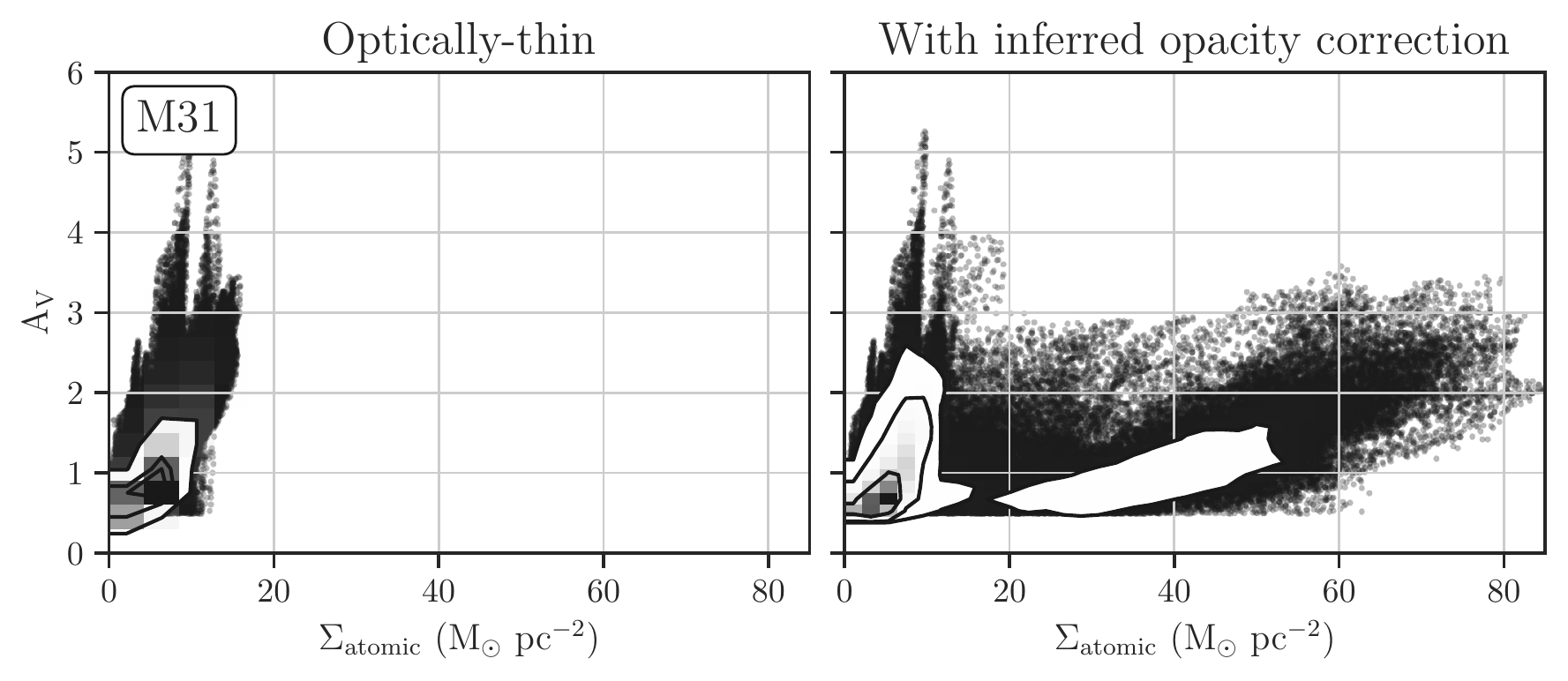}
\caption{\label{fig:m31_av_hi}
Atomic gas surface density versus mean visual extinction $A_{\rm V}$ from \citet{DalcantonFouesneau2015ApJ...814....3D} without (left) and with (right) \hi opacity corrections from Equation~\eqref{eq:thickhi_profile}.
The data are convolved to a common resolution of $\sim60$~pc, matching the original $18\arcsec$ resolution of our \hi data for M31 (\S\ref{sec:observations}).
Similar to the dust surface density from IR emission (Figure~\ref{fig:dust_hi_gdr}), we find a roughly linear relation assuming optically thin \hi emission and a large deviation when the inferred \hi opacity corrections from the opaque model (Equation \ref{eq:thickhi_profile}) are included.
To be consistent with the inferred opacity corrections, we expect to find compact regions with large $A_{\rm V}$.
The lack of such features is consistent with the opaque model overestimating opaque \hi corrections and previous work \citep{WhitworthMarsh2019MNRAS.489.5436W}.
}
\end{figure*}

Figure~\ref{fig:m31_av_hi} shows the relation between $\Sigma_{\rm atomic}$ and $A_{\rm V}$ with and without the inferred \hi opacity correction from the opaque model (Equation \ref{eq:thickhi_profile}).
Similar to Figure~\ref{fig:dust_hi_gdr}, the inferred \hi opacity correction increases the scatter in the relation of $\Sigma_{\rm atomic}$ and $A_{\rm V}$.
In particular, lines-of-sight with large opaque \hi corrections require a population where $A_{\rm V}$ remains constant but $\Sigma_{\rm atomic}$ varies by a factor of $4$.
This difference appears hard to explain, since the inclusion of other ISM components (e.g., ionized or molecular gas) that are not accounted for create a larger discrepancy.

One potential explanation for this inferred high $\Sigma_{\rm atomic}$ at low $A_{\rm V}$ is if the dust emission is primarily from compact and unresolved sources.
However, these features would be hard to explain given the dust surface density remains low in Figure~\ref{fig:dust_hi_gdr}.
This is further disfavoured based on a recent comparison of dust emission and extinction on $\sim30$~pc scales by \citet{WhitworthMarsh2019MNRAS.489.5436W}.
They find that dust emission from compact sources would require far wider distributions in $A_{\rm V}$ than those found in \citet{DalcantonFouesneau2015ApJ...814....3D}.
The nature of these ``compact sources'' and their effect on the relation between atomic gas and dust would need to explain these apparent contradictions in the observed relations.

The comparison of atomic gas surface densities with and without \hi opacity correction with both dust emission and dust extinction maps brings us to the conclusion that the opaque \hi corrections inferred from the opaque model fits (Equation \ref{eq:darkhi}) are likely overestimated or highly uncertain.
Because of this, the scatter is significantly increased in the relation between $\Sigma_{\rm atomic}$ and $\Sigma_{\rm dust}$ when the inferred opaque \hi corrections are included, which is the opposite trend expected for significant opacity corrections.
These results qualitatively agree with the lack of evidence for the opaque \hi model fits based on the spectral fitting and model selection in \S\ref{sec:mod_comp}.

\section{Discussion}
\label{sec:discussion}

Our results suggest a vastly different picture than the ubiquity of opaque \hi in M31, M33, and the LMC proposed by \citet{Braun2009ApJ...695..937B} and \citet{Braun2012ApJ...749...87B}.
Here, we discuss reasons for this discrepancy (\S\ref{sub:discrepancies}), explore differences in the opaque \hi mass correction factors that we determine (\S\ref{sub:corrfactor}), and compare the opaque model fits with \tspin values derived from \hi absorption studies (\S\ref{sub:hi_absorp_studies})
We also state shortcomings of the models we fit (\S\ref{sub:model_limits}) and identify additional approaches to find opaque \hi using emission spectra (\S\ref{sub:next_steps}).

\subsection{Discrepancies with the \texorpdfstring{\citeauthor{Braun2009ApJ...695..937B}}{Braun et al.{}} interpretation}
\label{sub:discrepancies}

Our results are clearly discrepant from those in \citet{Braun2009ApJ...695..937B} and \citet{Braun2012ApJ...749...87B}.
These discrepancies result from (i) improved spectral resolution and a higher sensitivity on smaller spatial scales, and (ii) improvements in the modelling methodology.

Table~\ref{tab:observe_comp} summarizes the differences in our observations versus those from \citet{Braun2009ApJ...695..937B} and \citet[][]{Braun2012ApJ...749...87B}.
Our observations have a similar sensitivity at moderately higher spatial resolution but with a finer spectral resolution.
In \S\ref{sub:specres_effect_on_modelpref}, we demonstrate how coarser spectral resolution ($2.4$~\kms) leads to a larger $\Delta {\rm BIC}$ when comparing the fits of the opaque and multi-Gaussian fits, though we find that the latter is still preferred in the majority of the spectra ($\Delta {\rm BIC} < 0$).
This comparison shows the importance of spectral resolution for resolving blended spectral components that may otherwise appears as a ``top-hat'' like shape.

Another source for discrepancy in our results stems from the improvements in methodology used for fitting.
\citet{Braun2009ApJ...695..937B} do not explicitly model multiple components, and additional residuals after fitting Equation \eqref{eq:thickhi_profile} are assumed to be optically thin emission.
Further, the model selection criteria they use to distinguish between multiple components versus a flat-top is more coarse than our comparison.
They define spectra with reduced chi-square values of $\chi_{\rm r}^2 < 25$, based on visual inspection, to be a valid opaque \hi fit.
However, given there are faint features not modelled in their fits (see their figure~14), it is not clear whether this threshold is sufficient to distinguish between multiple components and faint emission not included in the fit.
\citet{Braun2012ApJ...749...87B} reject $4\%$ of their fits in M31 and $<1\%$ in M33 based on this threshold.

For comparison, if we were to apply the same restriction of $\chi_{\rm r}^2 < 25$ to our opaque model fits then we would find that \emph{none} of the opaque model fits to be excluded.
This difference is likely the result of different systematics in the observations, combined with the effect of spectral resolution as mentioned above.
This demonstrates that an absolute fit statistic value, like $\chi_{\rm r}^2$, produces different effects when applied to different data sets.
This makes it difficult to directly compare the fit quality.

\subsection{Correction factor for optically thick \texorpdfstring{\hibold}{HI} column density}
\label{sub:corrfactor}

The key outcome from previous studies using the opaque model is the ability to account for ``dark'' \hi over a large spatial area, overcoming limitations of other methods that require \hi absorption against background sources.
With a constraint on the opacity, the cold opaque atomic gas content can be estimated, yielding a mass correction factor for the atomic ISM relative to the optically thin mass.
In \citet{Braun2009ApJ...695..937B} and \citet{Braun2012ApJ...749...87B}, they infer that opaque \hi accounts for an additional $\sim35\%$ of atomic ISM mass in M31 and M33.
When using the opaque model (Equation \ref{eq:thickhi_profile}), we calculate similar inferred opaque correction factors from the dark \hi intensity (Equation~\eqref{eq:darkhi}) and the optically thin atomic mass conversion described in \S\ref{sec:dustcomparisons}.
However, in this section, we expand these estimates to demonstrate the effect of (i) model selection criteria and (ii) the inherent uncertainty of the opaque model fits.

When applying different choices of the model selection criteria, we find that the dark \hi mass estimate varies drastically.
To show this, we calculate the inferred mass correction factors from the opaque model when (i) no model selection is applied (i.e., all opaque model fits), (iii) fits where the inferred peak opacity is $\tau_{\rm p} < 5$ (i.e., to test how sensitive the mass correction factor is to extreme, but allowed, highly opaque model fits), and (iii) where the opaque model is preferred (i.e., fits with $\Delta {\rm BIC}>0$).
Table~\ref{tab:darkhimasslimits} shows these mass correction factors for the different selections and the fraction of the fit spectra included.
In all three cases, we find that the mass correction factors have large uncertainties.

First, we calculate the mass correction factors with no model selection ($\Delta {\rm BIC}$) applied, and so all of the opaque fits are included.
We further note that this is the closest comparison to the mass correction factors computed \citet{Braun2009ApJ...695..937B} and \citet{Braun2012ApJ...749...87B}, since their restriction of $\chi^2_{\rm r}<25$ removes none of our fits (\S\ref{sub:discrepancies}).
For M33, we recover a correction factor of $21\pm15\%$, which is smaller than the $\sim36\%$ found by \citet{Braun2012ApJ...749...87B}, but is consistent within the large uncertainty.
For M31, we find a much larger correction factor of $118\pm47\%$ compared to M33.
Despite the large uncertainty, this correction factor is not consistent within the $1\sigma$ uncertainty to the $30\%$ factor found by \citet{Braun2009ApJ...695..937B}.
The reason for finding such a large correction factor is due to including opaque fits to Equation \ref{eq:thickhi_profile} with very large ($\tau_{\rm p}>5$) inferred peak opacities, which correspond to large dark \hi masses that often exceed the corresponding optically thin \hi mass.
These lines-of-sight are evident in Figures~\ref{fig:delta_bic_vs_darknhi} and~\ref{fig:delta_bic_vs_darknhi_taulimonly} as the contours with large negative $\Delta {\rm BIC}$ values with $I_{\rm dark}\sim10^4$~\kks.
The preference for $\Delta {\rm BIC} \ll -10$ indicates that the fits to the opaque model are poor.
In particular, we note that these are likely the lines-of-sight that were excluded based on the $\chi^2_{\rm r}<25$ restriction used in \citet{Braun2009ApJ...695..937B} and \citet{Braun2012ApJ...749...87B}.
Here, however, we do not find that this criterion removes these cases of poor fits, demonstrating that an absolute threshold for a fit statistic is dependent on the data that are fit.

While we expect these lines-of-sight with a large inferred opacity from Equation \ref{eq:thickhi_profile} to be poor fits, strictly speaking, these large $\tau$ values are acceptable in the model. 
Additionally, Milky Way studies towards extreme star-forming regions, like W43, do infer lines-of-sight with a $240\%$ increase in atomic gas mass compared to the optically thin mass \citep{BihrBeuther2015A&A...580A.112B}.
However, such large correction factors seem to be rare in the Milky Way \citep{WangBeuther2020A&A...634A..83W} and may result from many cold and warm atomic gas layers towards these lines-of-sight through the Galactic plane.

\begin{figure*}
\includegraphics[width=\textwidth]{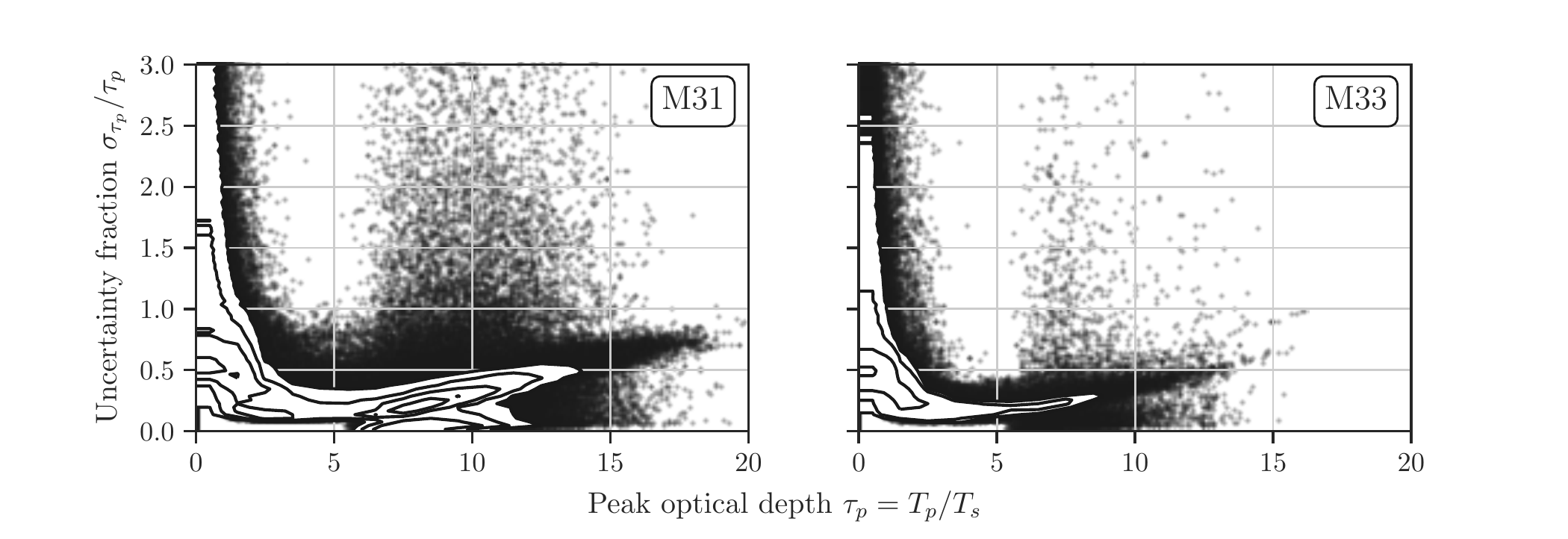}
\vspace*{-10px}
\caption{
\label{fig:m31_m33_peaktau_error}
Peak optical depth ($\tau_{\rm p} = T_{\rm p} / T_{\rm s}$) versus its uncertainty fraction ($\sigma_{\tau_{\rm p}} / \tau_{\rm p}$) for M31 (right) and M33 (left).
Contours and outliers levels are defined in Figure~\ref{fig:delta_bic_vs_darknhi}.
In the optically thin limit, the uncertainty on $\tau_{\rm p}$ is large as \tspin is unconstrained.
However, the uncertainty also becomes larger for $\tau_{\rm p} > 5$, particularly in M31.
Large uncertainties at for large $\tau_{\rm p}$ produce large uncertainties in the inferred dark \hi mass based on the opaque model fits (Equation \ref{eq:thickhi_profile}).
}
\end{figure*}

We control for the possibility of very large $\tau_{\rm p}$ being poor opaque model fits by excluding fits with $\tau_{\rm p}>5$.
This threshold is based on Figure~\ref{fig:m31_m33_peaktau_error}, where we show that the uncertainty increases for opaque model fits with approximately\footnote{The exact threshold on $\tau_{\rm p}$ is not critical as we impose this cut to demonstrate how it affects the mass correction factor.} $\tau_{\rm p}>5$.
Using this threshold, we compute mass correction factors of $11\pm14\%$ for M31 and $11\pm11\%$ for M33.
We note that negative correction factors are not defined based on Equation \ref{eq:darkhi} and that the large uncertainty range only reflects the break-down of our assumption of normally-distributed uncertainties near a mass correction $0\%$, or the optically-thin limit.
Removing these likely poor fits with large $\tau_{\rm p}$ significantly decreases the correction factors, however, large uncertainty ranges persist.

In the final comparison, we use the results from our model comparison test in \S\ref{sec:mod_comp} to only include fits where the opaque model is preferred ($\Delta {\rm BIC} > 0$).
Because the opaque model is preferred in only $\sim1\%$ of the spectra in both galaxies, the corresponding mass correction factors are also small.
We find correction factors of $1.3\pm0.7\%$ for M31 and $0.5\pm0.2\%$ for M33.

Together, these three approaches for calculating the opaque mass correction factor demonstrate its sensitivity to (i) the parameter space that is allowed for the model, and (ii) the ``goodness-of-fit'' criteria.
These factors make it difficult to robustly determine the opaque mass correction factor from fits to the single opaque component model (Equation~\eqref{eq:thickhi_profile}).

\begin{table*}
    \centering

\begin{tabular}{llll}
                             &   Sample  & M31 & M33 \\
                             & Selection &  &        \\\hline
Opt.-Thin Atomic Mass$^{\star}$ (\msol)    &    --                   & $1.7\times10^{9}$  & $1.8\times10^{9}$ \\[6px]  

Dark Atomic Mass$^{\star}$ (\msol)      & All                     & {$(2.0\pm0.8)\times10^{9}$} & {$(3.7\pm2.7)\times10^{8}$} \\
\% of Spectra                           &                         & $100\%$            & $100\%$           \\
\% Mass Difference                      &                         & {$118\pm47\%$}            & {$21\pm15\%$}           \\[6px]

Dark Atomic Mass$^{\star}$ (\msol)    & $\tau_{\rm p}<5$          & {$(1.8\pm2.3)\times10^8$ $^\dagger$} & {$(1.8\pm1.9)\times10^{8}$ $^\dagger$} \\
\% of Spectra                           &                         & $85.7\%$                      & $97.3\%$           \\
\% Mass Difference                      &                         & {$11\pm14\%^{\dagger}$}             & {$11\pm11\%^{\dagger}$}           \\[6px]

Dark Atomic Mass$^{\star}$ (\msol) & $\Delta {\rm BIC} > 0$  & {$(2.1\pm1.2)\times10^{7}$}  & {$(8.8\pm3.4)\times10^{6}$} \\
\% of Spectra                      &                              & $1.6\%$                     & $0.8\%$           \\
\% Mass Difference                      &                         & {$1.3\pm0.7\%$}      & {$0.5\pm0.2\%$}           \\

\end{tabular}

    $^{\star}$ Including $1.36$ factor for He and metals.
    \caption{Dark atomic ISM mass in M31 and M33 estimated with (i) no selection criteria, (ii) only spectra with peak optical depths of $\tau_{\rm p} < 5$ based on the opaque model, and (iii) spectra where the opaque model is preferred ($\Delta {\rm BIC} > 0$; Equation~\eqref{eq:thickhi_profile}).
    We highlight that the dark atomic mass converted from $I_{\rm dark}$ (Equation \ref{eq:darkhi}) must be $>0$ and our assumption of normal-distributed uncertainties breaks down when the uncertainty is large, which we indicate in the table with a $^\dagger$.
    We note that all of our fits satisfy the $\chi^2_{\rm r} < 25$ criterion from \citet{Braun2009ApJ...695..937B}, and therefore their selection criterion is equivalent to using all of the opaque \hi fits.
    The table also shows the percent of spectra after the selection criterion is applied, and the percent mass difference compared to the optically thin mass.
    The optically thin atomic ISM mass is calculated from the integrated intensity maps excluding pixels without valid spectral fits.
    We find large differences in the mass correction factors that depend strongly on the sample selection, and similarly the goodness-of-fit.
    This strong dependence on sample selection suggests that the dark \hi mass from the opaque model (Equation~\eqref{eq:thickhi_profile}) is highly uncertain.
    }
    \label{tab:darkhimasslimits}
\end{table*}

In each of the cases presented above, the uncertainty of opaque mass correction factor is large.
This uncertainty results from the non-linear relation between the opaque \hi intensity $I_{\rm dark}$ on $\tau_{\rm p}=T_{\rm peak} / T_{\rm spin}$.
The fit parameters \tpeak and \tspin have appreciable uncertainties from the fits, and this is reflected in the uncertainty on $\tau_{\rm p}$ (see Figure~\ref{fig:m31_m33_peaktau_error}).
The large uncertainty on $I_{\rm dark}$, and consequently on deriving the dark atomic mass, adds to the difficulty in robustly measuring the opaque mass correction factor.
We note that this uncertainty was not explored in \citet{Braun2009ApJ...695..937B} and \citet{Braun2012ApJ...749...87B}.

Finally, we highlight that finding a $\sim1\%$ correction factor does \emph{not} imply a lack of opaque or cold \hi in either M31 and M33.
Our results only demonstrate that \hi spectra are not well-described by a single opaque \hi component.
We discuss the limitations to this model and our comparisons further in \S\ref{sub:model_limits}.

\subsection{Comparisons with \hibold absorption studies of M31 and M33}
\label{sub:hi_absorp_studies}

Observations of \hi absorption provide strong constraints on the \hi opacity that can be compared with the single component opaque model fit to the \hi emission.
\citet{DickeyBrinks1993ApJ...405..153D} present \hi absorption measurements against 11 and 10 background continuum sources ($>5$~mJy) for M31 and M33, respectively.
By matching the measured absorption to the \hi emission on larger scales, they infer the fraction of \hi mass in the CNM\footnote{\citet{DickeyBrinks1993ApJ...405..153D} assume that $T_{\rm spin}=60$~K. Lowering \tspin will decrease $f_{\rm CNM}$.}, $f_{\rm CNM}$, along these lines-of-sight and find average values of $f_{\rm CNM}\approx0.4$ in M31 and $0.15$ in M33.
Though measured towards only a small number of lines-of-sight, these $f_{\rm CNM}$ values fall within a similar range to those measured in the Milky Way \citep[$f_{\rm CNM}\approx0.3$;][]{StanimirovicMurray2014ApJ...793..132S,MurrayStanimirovic2018ApJS..238...14M}, LMC \citep[$f_{\rm CNM}\approx0.3$;][]{Marx-ZimmerHerbstmeier2000A&A...354..787M}, and SMC \citep[$f_{\rm CNM}\approx0.2$;][]{JamesonMcClure-Griffiths2019ApJS..244....7J}.

\begin{figure}
\includegraphics[width=0.45\textwidth]{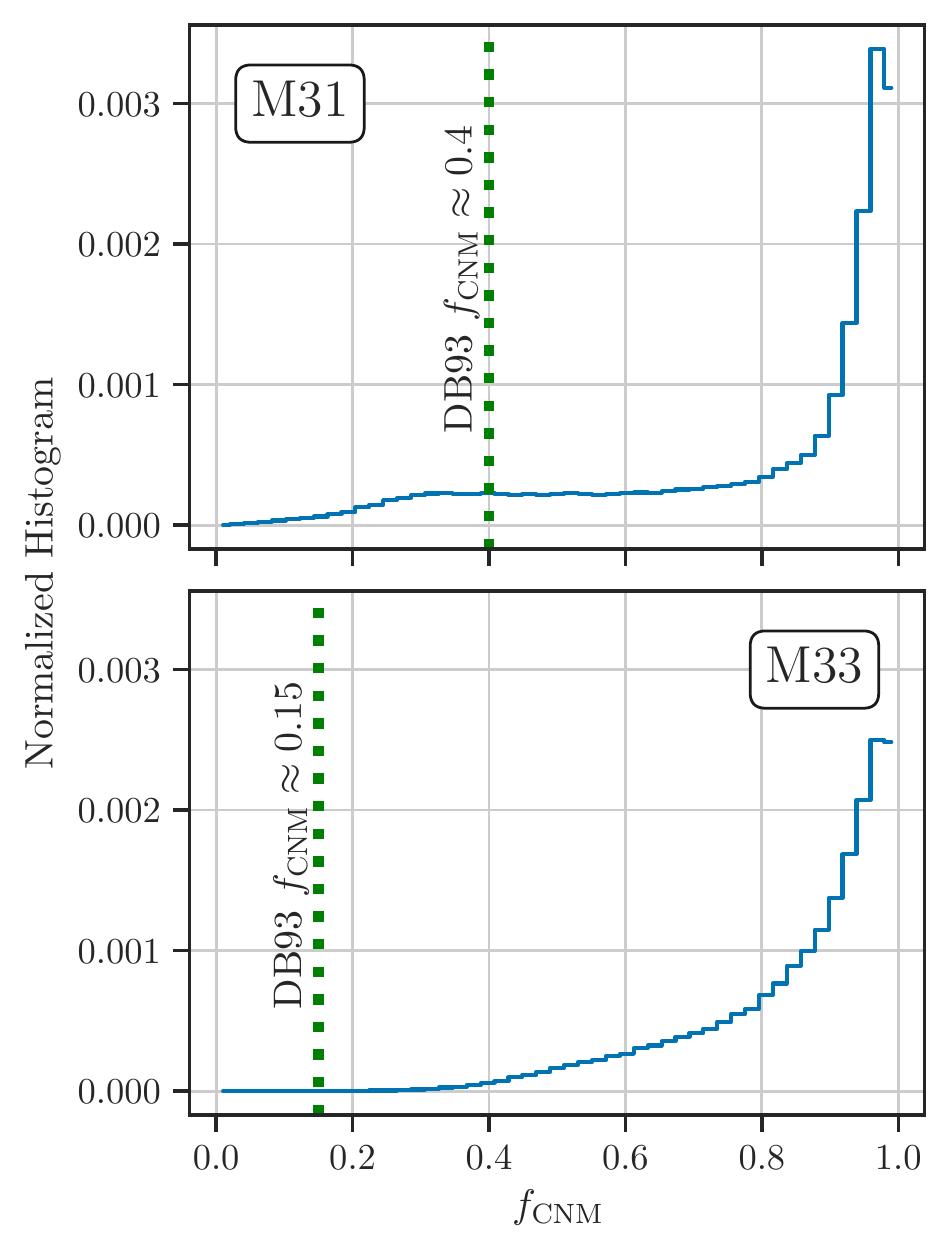}
\vspace*{-10px}
\caption{
\label{fig:m31_m33_fCNM_hists}
Histograms of $f_{\rm CNM}$ measured from the single opaque component model where the fits are consistent with CNM temperatures ($T_{\rm spin} < 250$~K; Equation \ref{eq:fCNM}).
The average $f_{\rm CNM}$ derived from \hi absorption studies by \citet[][DB93]{DickeyBrinks1993ApJ...405..153D} are shown with the vertical green dotted lines.
Though the \hi absorption measurements are based on a small number of lines-of-sight, the single opaque component model infers much higher $f_{\rm CNM}$.
In particular, the histograms peak near $f_{\rm CNM}\sim1$, which would imply a population of lines-of-sight with \emph{only} CNM, inconsistent with \hi absorption studies of the Milky Way and other Local Group galaxies.
}
\end{figure}

To compare with $f_{\rm CNM}$ values from \hi absorption studies, we use the opaque model fits to estimate $f_{\rm CNM}$ where the inferred spin temperature is constrained and implies the presence of CNM \citep[$T_{\rm spin} < 250$~K;][]{MurrayStanimirovic2018ApJS..238...14M}.
Following \citet{Braun2009ApJ...695..937B}, we compute the effective total \hi intensity from the opaque model including the ``dark'' component ($I_{\rm op.\ model}$), plus the difference of the observed integrated intensity (i.e., the zeroth moment) minus the opaque model fit:
\begin{equation}
    \label{eq:totalint_opaque}
    I_{\rm total} = I_{\rm op.\ model} + \left( I_{\rm int} - \int T_{\rm s} \big[ 1 - {\rm exp}\left[-\tau(v) \right] \big] dv \right),
\end{equation}
where $I_{\rm int}$ is the zeroth moment integrated intensity, and the second term is assumed to be the optically-thin component.
Since we compare only the fits where $T_{\rm spin}<250$~K, $I_{\rm CNM} = I_{\rm op.\ model}$, and the CNM fraction can be estimated as:
\begin{equation}
    \label{eq:fCNM}
    f_{\rm CNM} = \frac{I_{\rm CNM}}{I_{\rm total}}.
\end{equation}
Figure \ref{fig:m31_m33_fCNM_hists} shows that the distributions of $f_{\rm CNM}$ from the opaque model tend to be quite large, with a peak near $f_{\rm CNM}\sim1$ in both galaxies.
Such large $f_{\rm CNM}$ values are difficult to reconcile with the $f_{\rm CNM}<0.4$ found from \hi absorption studies in the Milky Way and other Local Group galaxies, as discussed above.
However, the lines-of-sight with $f_{\rm CNM}\sim1$ correspond to those with the largest $\tau$, lowest \tspin, and largest $I_{\rm dark}$ (Equation \ref{eq:darkhi}).
As we show in \S\ref{sub:model_tests}, these lines-of-sight are also those with the \emph{strongest} preference for the multi-component Gaussian model (Figures \ref{fig:delta_bic_vs_darknhi} \& \ref{fig:delta_bic_vs_darknhi_taulimonly}).
Therefore, the discrepancy in $f_{\rm CNM}$ with \hi absorption studies is consistent with the single opaque model poorly respresenting the observed \hi spectra.

\subsection{Limitations of the \hibold models}
\label{sub:model_limits}

We compare two models to describe \hi in this work and each has its own limitations.
We describe these limitations here and how this limits making general conclusions about the \hi opacity in M31 and M33.

The opaque model that we use, similar to the model used in \citet{Braun2009ApJ...695..937B} and \citet{Braun2012ApJ...749...87B}, is primarily limited by having a single component.
The example spectra in Figure~\ref{fig:m31_m33_peaktau_error} demonstrate this limitation clearly as many show $>1$ distinct components.
This is what drives our model comparison results (\S\ref{sec:mod_comp}).
However, it does not test whether those individual components are better modelled by this opaque model.

The limitation in applying the opaque model to individual \hi components is that it is difficult to distinguish independent components versus those that are $>1$ Gaussians blended together \citep{RohlfsBraunsfurth1972AJ.....77..711R,PeekHeiles2011ApJ...735..129P}.
One way to overcome this limitation is to identify components that are most likely to have opaque \hi based on, for example, the \hi components associated with tracers of molecular gas.

The key limitation in the multi-component Gaussian model is the inherent uncertainty when components overlap because Gaussians do not form an orthogonal basis set.
This means that the properties of individual components from the multi-Gaussian fit models (Figure~\ref{fig:m31_m33_spectra_wfits}) should be treated carefully for physical interpretations \citep{RohlfsBraunsfurth1972AJ.....77..711R,Haud2000A&A...364...83H}.
Our fitting approach uses spatial similarities to improve spectral fitting, which encourages spatial coherence, but remains uncertain where blended components are correlated.
Improving how spatial information is included in the spectral fits can minimize some of this uncertainty \citep[][]{MarchalMiville-Deschenes2019A&A...626A.101M,RienerKainulainen2020A&A...633A..14R}.

There are additional \hi features that are also not accounted for in either model.
For example, we do not include the possibility for \hi self-absorption (HISA) features, which can be distinguished from the opaque \hi model as a narrow absorption ``dip'', e.g., as those frequently seen in Milky Way \hi spectra \citep{Gibson2005ApJ...626..195G}.
While self-absorption may be detected towards some lines-of-sight, we do not expect our observations to be sensitive to strong HISA features due to the $>60$~pc linear resolution.
In agreement with this, \citet{KochRosolowsky2019MNRAS.485.2324K} find the velocity at peak \hi and CO brightness temperature to be typically consistent within the $2.6$~\kms CO velocity channel size, rather than additional scatter or offsets in this relation if the CO peaked at a strong self-absorption feature.
In qualitative agreement with these results, \citet{LiuLi2019ApJ...887..242L} use CO components to derive \hi self-absorption in the LMC and find relatively small corrections on $15$~pc scales.

\subsection{Future directions for identifying opaque \hibold in nearby galaxies}
\label{sub:next_steps}

Our results show that very few lines-of-sight in M31 and M33 are well-described by a single opaque \hi component and that a multi-component model is strongly preferred.
This indicates that we are likely recovering a mixture of atomic ISM phases, in addition to multiple kinematic components reflecting galactic structure.
For example, our M33 \hi observations \citep{Koch2018MNRAS} detect a lagging thick \hi disc in M33 \citep[see also][]{Kam2017AJ....154...41K}.
Disentangling this spectral complexity and linking it to the underlying atomic ISM structure is an obvious next step in exploring these observations.
One crucial link in this next step will be to determine whether components of the \hi emission spectra \emph{can} identify cold opaque \hi, thereby providing an improved estimate of the ``dark'' opaque \hi mass.

With current observations, we suggest three approaches to further search for opaque \hi without requiring \hi absorption observations against bright background sources:
\begin{enumerate}
    \item The line width of \hi components can distinguish between emission from different atomic ISM phases \citep[e.g.,][]{Haud2000A&A...364...83H,MarchalMiville-Deschenes2019A&A...626A.101M}.
    However, the degeneracy of line broadening from turbulence limits line width measurements to an upper limit of the \hi temperature.
    In the simplest case of a constant WNM temperature of $6000$~K, measured line widths $<6$~\kms must be non-WNM.
    For much narrower line widths ($\leq2.5$~\kms), the equivalent temperature is $<1000$~K without including turbulent broadening, making it more likely that these features are indeed from the CNM.
    We identify few distinct narrow components from the multi-component Gaussian fits, however, there remains ambiguity about the completeness of these narrow components given the issues with blended components (see Appendix~\ref{appsub:mgauss_distinct_blended}).
    This issue will be explored in E.~W.~Koch et al. (in preparation).
    
    Resolving finer spatial scales will also assist in resolving some of these spectral blending issues, since the blending will be minimized when the associated spatial extent of these components becomes resolved.
    We have in-hand additional VLA \hi observations resolving $10\arcsec \approx 35$~pc scales in M31, and $8\arcsec \approx 30$~pc in M33 that we will use to explore this avenue (E.~W.~Koch et al. in preparation).
    
    \item The degeneracy between spectral components is a key limitation of the multi-component Gaussian modelling.
    As we note in \S\ref{sub:model_limits}, using constraints from other spectral lines can assist in resolving between this degeneracy by restricting where we search for signatures of opaque \hi.
    Additionally, these comparisons could lead to HISA detections \citep[e.g.,][]{LiuLi2019ApJ...887..242L,WangBihr2020A&A...634A.139W}.
    
    \item ``Dark'' neutral ISM mass can further be constrained by combining dust, atomic ISM and molecular ISM tracers.
    Similar comparisons have been made in both M31 and M33 \citep{LeroyBolatto2011ApJ...737...12L,DalcantonFouesneau2015ApJ...814....3D,GratierBraine2017A&A...600A..27G}, as well as nearby galaxies \citep{SandstromLeroy2013ApJ...777....5S}, though typically on $>100$~pc scales.
    Extending similar comparisons to $\sim30$~pc scales will be helpful for resolving between \htwo-dominated lines-of-sight towards GMCs and the surrounding atomic gas, which may include a cold, opaque component \citep[e.g., see][and Smith et al. in preparation for IR dust observations on these scales in M33 and M31 ]{WilliamsGear2019MNRAS.483.5135W}.
    
    One issue in using the dust-\hi-CO comparison will be distinguishing whether the ``dark'' ISM component is opaque \hi or CO-dark \htwo \citep[e.g., see][for constraints on this distinction in the Milky Way]{MurrayPeek2018ApJ...862..131M}.
    However, this comparison on $\sim30$~pc scales will provide strong constraints on the \emph{total} dark gas mass, which has otherwise been limited to statistical arguments over large spatial areas \citep[e.g.,][]{GratierBraine2017A&A...600A..27G}.
    The \hi line shape and possibility of individual opaque components, as described above, may be the key for making this distinction.
    Additional observations of \cii and/or \ci can help resolve this degeneracy in the phase of a ``dark'' component \citep{BisbasSchruba2019MNRAS.485.3097B,MaddenCormier2020A&A...643A.141M}, however, mapping the large angular areas covered by M31 and M33 at high resolution remains prohibitive for current sub-mm and IR telescopes.

\end{enumerate}

While the approaches described above seem hopeful, there remains the clear advantage of a direct measurement of \hi opacity from the absorption against bright background sources, despite the limited number of lines-of-sight where the absorption can be measured \citep{DickeyBrinks1993ApJ...405..153D}.
Though spatially limited, these \hi absorption measurements are crucial to compare with opacity measurements from other methods (e.g., Figure \ref{fig:m31_m33_fCNM_hists}).

\section{Summary}
\label{sec:summary}

In this paper, we explore whether extragalactic \hi emission can be described by a single opaque velocity component model to constrain the \hi line opacity and measure the ``missing'' opaque atomic ISM mass.
Using new sensitive 21-cm \hi observations at high spectral resolution in M31 and M33, we compare two common spectral models: a multi-Gaussian model and a single-component, opaque \hi model \citep[e.g.,][]{Braun2009ApJ...695..937B}.
With the improved spectral resolution of our data, we demonstrate a strong preference for the multi-Gaussian model for $>80\%$ of the spectra, while $<2\%$ prefer the opaque model and the remainder does not have a strong model preference.
Adopting the opaque model, we further compare the inferred \hi opacity corrections to dust emission and dust extinction properties, where we find that the \hi opacity corrections require large variations in dust abundance that lack observational evidence.
Combining our model selection and dust comparison results, we show that models for a single opaque \hi component are not supported by the observations, which demonstrates there is a lack of lines-of-sight dominated by \emph{only} cold and opaque \hi on $\sim80$~pc scales.
Finally, we show that the inferred optical depth from the opaque model fit to \hi emission alone is subject to significant systematic uncertainty, which in turn, leads to large uncertainties on the inferred opaque \hi mass.
Our main conclusions from this work are:

\begin{enumerate}
    \item Improvements in the spectral resolution of extragalactic 21-cm \hi data reveal a wealth of spectral structure.
    The observations we present extend detailed \hi studies that were previously restricted to the Milky Way and the Magellanic Clouds.
    
    \item We identify a strong preference for an optically-thin, multi-Gaussian model over a single opaque velocity component across both M31 and M33 by comparing the difference of the Bayesian Information Criteria (BIC) between the two models (Figure~\ref{fig:m31_m33_delta_bic_map}).
    This preference persists when considering only opaque models with a large inferred optical depth ($\tau_{\rm p} > 1$), and when the BIC is computed only over velocities where the opaque model measures $\tau(v) > 0.5$.
    
    \item We verify our model selection test by applying the same fitting procedures to $20{,}000$ spectra drawn from mock observations of the opaque model, sampled at the same $0.42$~\kms channel width and noise levels as the observations.
    We find that $<1.5\%$ prefer the multi-Gaussian model, which is the opposite trend the we find from the observations.
    This test demonstrates that our observations and analysis methods are sensitive enough to distinguish between the two models.
    
    \item The inferred opaque \hi correction factors produce highly uncertain estimates of the opaque \hi mass.
    Using the opaque correction factor from all lines-of-sight, we find mass correction factors of $118\pm47\%$ for M31 and $21\pm15\%$ for M33, where the ranges are $1\sigma$ uncertainties.
    However, these mass correction factors reduce to just $\sim1\%$ in both galaxies when we restrict to only using lines-of-sight where the single opaque component model is preferred.
    This large range, coupled with the large uncertainty, demonstrates that the inferred opaque \hi mass factors are highly sensitive to the goodness-of-fit threshold and the allowed parameter range for the fit.

    \item We show that the scatter in the relation between $\Sigma_{\rm atomic}$ and $\Sigma_{\rm dust}$ is significantly increased when the inferred \hi opacity corrections are included (Figures~\ref{fig:dust_hi_gdr} and~\ref{fig:m31_av_hi}).
    This is the opposite of what is expected if the \hi surface density is missing a component from the optically thin assumption relative to the dust, tracing the total neutral gas surface density.
    This increased scatter from the inferred \hi opacity correction can only be explained for a population of lines-of-sight with large gas-to-dust ratios (${\rm GDR}\approx1000$) and $A_{\rm V} < 2$~mag over a range of $\Sigma_{\rm atomic}\approx20{-}80$~\msolpcsq, which lacks observational evidence.
    Since the scatter is reflected in $A_{\rm V}$ and the dust surface density from IR SED fitting, this rules out the possibility of cold dust not accounted for in the far-IR.

\end{enumerate}

Our results highlight that estimates of the opaque \hi mass from the opaque spectral line model are not reliable on 100~pc scales.
The large difference in the inferred dark atomic \hi mass based on fits to the opaque model point at the critical role of model assumptions on interpretations of the atomic ISM.
We describe how future studies may improve on this modelling.
However, given our current results, we suggest that \hi opacity is best constrained by comparing \hi absorption to background sources with nearby emission \citep[e.g.,][]{DickeyMcClure-Griffiths2003ApJ...585..801D}, or through \hi self-absorption \citep[e.g.,][]{Gibson2005ApJ...626..195G,WangBihr2020A&A...634A.139W}, though current instrumentation limits the use of these methods in M31 and M33.

\section*{Acknowledgments}

We thank the reviewer for insightful comments that improved the clarity of this work.
EWK and EWR are supported by a Discovery Grant from NSERC (RGPIN-2017-03987).
EWK acknowledges support in completing this paper from the Smithsonian Institution as a Submillimeter Array (SMA) Fellow.
TGW acknowledges funding from the European Research Council (ERC) under the European Union’s Horizon 2020 research and innovation programme (grant agreement No. 694343).
The work of J.C., K.S., I.C., A.K.L., and D.U. is supported by NASA ADAP grants NNX16AF48G and NNX17AF39G and National Science Foundation grant No. 1615728. The National Radio Astronomy Observatory and the Green Bank Observatory are facilities of the National Science Foundation operated under cooperative agreement by Associated Universities, Inc.
This research was enabled in part by support provided by WestGrid (www.westgrid.ca), Compute Canada (www.computecanada.ca), and CANFAR (\url{www.canfar.net}).

\textbf{Code Bibliography: }
CASA \citep[versions 4.4, 4.7, 5.4;][]{casa_mcmullin2007} --- astropy \citep{astropy,astropy2} --- radio-astro-tools (spectral-cube \citep{SPECTRALCUBE2020}, radio-beam, uvcombine; \url{radio-astro-tools.github.io}) --- matplotlib \citep{mpl} ---  seaborn \citep{seaborn} --- lmfit \citep{lmfit} --- numpy \& scipy \citep{2020SciPy-NMeth} --- uncertainties \citep{uncertainties}\\

\section*{Data Availability}

The software and code used in this paper are publicly available.
Scripts to reproduce the results in the paper are available at: \url{https://github.com/e-koch/HI-LineShapes}\footnote{Code DOI: \url{https://doi.org/10.5281/zenodo.4665936}}
The \hi data products for M33 were released with \citet{Koch2018MNRAS}.
The \hi data products for M31 will be publicly released upon acceptance of in preparation manuscripts.

\bibliographystyle{mnras}
\bibliography{ref}

\appendix

\section{Imaging approach and Single-dish combination} 
\label{app:imaging_approach}

In this section, we describe the $uv$-data handling and imaging steps for the M31 VLA observations (\S\ref{sub:vla}) in further detail.
The procedure is similar to that used in \citet{Koch2018MNRAS} for the M33 VLA observations, with some difference unique to M31 as a target and alterations to the imaging steps.

The M31 data from individual observing sessions were transformed into the LSRK frame and observations of matched VLA configurations were concatenated.
We then subtracted continuum sources from the data in the $uv$-plane using \textsc{uvcontsub} with a zeroth-order spectral fit.
The zeroth-order constant fit does not leave noticeable residuals in the continuum-subtracted data.

Next, we combined observations taken in different configurations.
Because we imaged the data at the native spectral resolution, we paid special attention to matching the spectral channel edges between the data for the 3 configurations.
By default (as of CASA v5.6 and before), regridding over shifts smaller than the channel size in \textsc{mstransform} will use linear interpolation that creates a ``beat'' pattern in the outputted data.
While the ``fftshift'' method for regridding should handle this, we found that the implementation in CASA v5.4.1 created large deviations near the edges of the spectral range in the outputted data.
These issues are discussed at length in \citet{Leroy20}.

For our M31 observations, the difference in the spectral channel edges are very similar across the different configurations ($\ll1\%$ the channel width).
Because of this, we split the channels according to the velocity range that will be imaged to the nearest matching channel number and concatenated these channel-matched data sets together.
A small frequency tolerance was allowed in the concatenation.

To handle the large computational time needed for imaging the \hi data, we imaged and deconvolved each spectral channel separately, similar to \citet{Koch2018MNRAS}.
Though this includes the additional step of splitting the $uv$-data, this approach minimizes the size of the $uv$-data and the time for input/output operations.
Furthermore, each channel can be imaged independently and can, for example, be parallelized by deploying the imaging for each channel as separate jobs on a cluster.
This approach has two benefits: (i) different combinations of imaging parameters can be rapidly explored on a small number of channels; and (ii) divergent deconvolution solutions can be corrected without the need to re-image other spectral channels.
We split the \hi uv-data into individual channels at the target spectral resolution---in this case, we use the native spectral resolution of the data, $0.42$~\kms.

We imaged each spectral channel in two steps.
First, we deconvolved the data to the $2\sigma_{\rm rms}$ level within a pre-defined mask.
An outer $uv$-taper of $5\arcsec$ was applied to produce a $\sim18\arcsec$ beam resolution, roughly equivalent to VLA's C-configuration.
Since the \hi emission is highly extended, we used multi-scale CLEAN \citep{Cornwell2008ISTSP...2..793C} with scales of $[0,18,36,72,180,360,720,1440]\arcsec$, spanning the $uv$-coverage of the VLA observations\footnote{Different choices of scales for the multi-scale CLEAN did not clearly change the final restored images.}, and major cycles were triggered every $500$~iterations, which assisted the convergence of the deconvolution.
We produced signal masks from the previously-imaged D-configuration ($58\arcsec$) mosaic covering the entire star-forming disc\footnote{To be presented in E.~W.\ Koch et al. (in preparation).} using the masking procedure described in \citet{Koch2018MNRAS} and also applied to the M33 \hi data cube.
Since the beam size differs by only a factor of $\sim3$ between the D-configuration data and the data product derived here, the D-configuration mosaic provides excellent prior information at higher resolution in M31.
Using these signal masks was critical due to the highly elongated and extended \hi emission for the blue-shifted channels in this portion of M31.
Without the mask, there was a tendency for the deconvolution to diverge because of the large negative bowls between the blue-shifted ``limbs,'' which were difficult to account for with the symmetric multi-scale CLEAN elements.

After the first stage, we noticed low-lying emission or mild artefacts to remain in the residual and restored images.
To handle this, we continued in the deconvolution in a second step by removing the signal mask from the first step.
Since the vast majority of the bright emission is deconvolved in the first step, there is no longer an issue with negative bowling causing a divergence.
We continue to deconvolve until the residuals in the entire map reach the $2\sigma_{\rm rms}$ level.


All imaging was run on Compute Canada's Cedar cluster\footnote{\url{docs.computecanada.ca/wiki/Cedar}}.
Each channel was imaged in parallel using 8~cores and at least 4~GB of memory per core.
We imaged 4~channels at once on a single 32 core node with 128~GB of memory, the standard size available at the time on the Cedar cluster.
Each imaging stage required $20{-}30$~hr of wall time to complete per channel.

Once the imaging was completed, we recombined the channels to form the final spectral line data cube.

Finally, we included short-spacing information by feathering the VLA data cube with Effelsburg \hi observations from EBHIS \citep{WinkelKerp2016A&A...585A..41W} using the \textsc{uvcombine} package\footnote{\url{github.com/radio-astro-tools/uvcombine}}.
The EBHIS data have a $10\farcm8$ beam size that moderately overlaps with the shortest VLA baselines covering scales up to $16\arcmin$.
The overlapping range is less than the factor of $1.7$ recommended by \citet{Kurono2009PASJ...61..873K} for deriving a consistent scale factor in the overlap of the two data sets, however, we combine overlapping $uv$ scales over many spectral channels to increase the number of samples to improve statistical estimates.

To combine the two datasets, we spectrally oversample the EBHIS data cube from the original $1.3$~\kms to match the $0.42$~\kms resolution of the VLA data.
Because of the coarse resolution ($10\farcm8\approx2.6$~kpc), the \hi spectra with bright emission are smooth and upsampling by $\sim3\times$ does not affect the single dish spectral shape.
We note, however, that this assumption about the upsampling holds only at very high signal-to-noise.

To ensure the data are optimally combined, we performed two $uv$-overlap tests recommended by \citet{Stanimirovic1999PhDT} and also used for the M33 data \citep{Koch2018MNRAS}.
Both of these are implemented in the \textsc{uvcombine} package.
First, we test whether there is a nonzero slope between the ratio of $I_{\rm VLA}/ I_{\rm EBHIS}$ with the $uv$-distance when weighted by the single-dish beam, as is used in feathering.
If there is a nonzero slope, it suggests that the weighting from the single-dish data differs from the data, and the beam size moderately differs from what was assumed.
We find no trend with~$k$ that would be consistent with the reported $10\farcm8$ EBHIS beam size from \citet{WinkelKerp2016A&A...585A..41W}

Second, we calculated the typical ratio of $I_{\rm VLA}/ I_{\rm EBHIS}$ to determine whether the flux calibration agrees between the two data sets.
This test is best performed when the interferometric data cover most of the observed source, particularly for the large extended structure in M31 and the large EBHIS beam size.
Because of this, we did this test using the full VLA D-configuration mosaic rather than the smaller $18\arcsec$ mosaic used throughout this paper\footnote{When using a smaller mosaic, the single-dish observations are tapered by the interferometric primary beam response, and this test depends more strongly on the low level behaviour of the response function.}.
Using the full mosaic improves the ratio test uncertainty as it includes more signal from the source, as well.

We take the Fourier transform of 500~channels in both the VLA and EBHIS data and sample the angular scales from $10.8\mbox{16}\arcsec$ that both the EBHIS and VLA observations are sensitive to.
We then combine these samples across channels assuming the flux scaling does not changes across a narrow spectral range.
We then model the histogram of $I_{\rm VLA}/ I_{\rm EBHIS}$ as Cauchy (or Lorentzian) distribution, following \citet{Koch2018MNRAS}, that accounts for high- and low-tails in the distribution from noise in both the numerator and denominator.
From the fitted Cauchy distribution, we find a relative scaling factor of $1.07\pm0.01$.
Since the ratio is $>1$, this scaling factor applied to the EBHIS data increased the flux by $10\%$.
In an independent test, \citet{BlagraveMartin2017ApJ...834..126B} find a similar scaling factor of $1.10\pm0.01$ was needed for combining their observations from the Dominion Radio Astrophysical Observatory Synthesis Telescope with short-spacing from EBHIS.

The final VLA+EBHIS M31 \hi data cube is the feathered product after applying a scaling factor of $1.07$ to the EBHIS data.
The data has a beam size of $18\arcsec\times16\arcsec$ and a sensitivity of $2.8$~K per $0.42$~\kms channel.
Within each channel, this corresponds to a $5\sigma_{\rm rms}$ optically thin \hi column density of $9.8\times10^{18}$~\cmtwo.
This sensitivity is roughly equivalent to the sensitivity of the high-resolution WSRT \hi observations from \citet{Braun2009ApJ...695..937B} in $2.3$~\kms spectral channels.

\section{Multi-Gaussian model testing}
\label{app:gaussian_model_testing}

In \S\ref{sub:gauss_model}, we describe our approach for modelling the \hi spectra with multiple Gaussian components, building on previous work by \citet{Lindner2015AJ....149..138L} and \citet{RienerKainulainen2020A&A...633A..14R}.
Here, we test this method on synthetic spectra drawn from randomly-selected multi-Gaussian models sampled at the $0.42$~\kms spectral resolution of our observations with similar noise levels added.

We produce multi-Gaussian synthetic spectra in three steps:
\begin{enumerate}
    \item We draw the number of Gaussian components randomly between 1 and 8, where the values are weighted based on the fraction of M31 spectra with a given number of Gaussian components (e.g., $\sim40$\% with 1 component, $\sim30$\% with 2 components, $\sim20\%$ with 3 components, and $\sim10\%$ with $>3$ components).
    
    \item Next, we randomly draw the amplitude, centroid, and line width from uniform distributions again matching the range of parameters recovered from the observations.
    Specifically, we sample parameters from uniform distributions:
    \begin{align}
        \label{eq:app_synth_distribs}
            A \, &\sim \, \mathcal{U}\left(2.7, 120 \right) \, {\rm K} \\
            v_0 \, &\sim \, \mathcal{U}\left(-50, 50\right) \, {\rm km \ s}^{-1}\\
            \sigma \, &\sim \, \mathcal{U}\left(0.1, 60 \right) \, {\rm km \ s}^{-1}. &&
    \end{align}  
    The lower limit for $A$ is the rms noise in the M31 data.
    We allow for narrow features below the $0.42$~\kms channel width to test the recovery of unresolved features where the averaging within channels biases the line shape \citep[e.g.,][]{KochRosolowsky2018RNAAS...2..220K}.

    While the observed parameter distributions show correlations between the amplitude and line width, we do not include these correlations for the synthetic spectra.
    This choice to draw from uncorrelated uniform distributions provides a simpler way to determine completeness limits for the observed parameter distributions.
    
    \item The synthetic spectrum is evaluated at $4\times$ finer spectral resolution compared to the observations.
    We then average this model spectrum to reach $0.42$~\kms channels, matching the observations.
    Finally, we add Gaussian noise sampling from $\mathcal{N}(0, 2.7)$ to match the M31 rms noise.
    
\end{enumerate}

We draw $10{,}000$ synthetic spectra to test our Gaussian fitting procedure.
To examine the affect of the noise alone, 10 different noise draws are added to each synthetic spectrum and fit.
Though not fully independent, this yields a total $100{,}000$ fits to examine the fitting method.
From fitting the synthetic spectra, we present two results: (1) the recovery fraction between the actual and fit number of Gaussian components, and (2) the parameter distributions of components correctly recovered in the fits.

Due to the issue of covariance between Gaussian components, we introduce a ``distinctness'' criterion to distinguish between Gaussian components that are blended (large covariance) versus distinct (small covariance).
Distinct components in a multi-Gaussian spectrum are characterized by a clear peak, corresponding to a local minimum in the 2nd derivative, which allows for that shape to be well-constrained when fitting.
Because the component shape is distinguished in the multi-Gaussian spectrum, we can identify distinct components using a similar approach to the ``derivative spectroscopy'' we use in our fitting method \citep[\S\ref{sub:gauss_model};][]{Lindner2015AJ....149..138L}.

\subsection{Classifying distinct and blended Gaussian components}
\label{appsub:mgauss_distinct_blended}

We define a component as ``distinct'' when the ratio between the 2nd derivative of the total model versus the single component exceeds $f_{\rm Distinct}=0.75$:
\begin{equation}
    \label{eq:distinct_def}
    f_{\rm Distinct} = \frac{\displaystyle\frac{{\rm d}^2 T_b (v_{0,i})}{{\rm d}v^2}}{\displaystyle\frac{{\rm d}^2 T_{b, i} (v_{0,i})}{{\rm d}v^2}}~,
\end{equation}
following the notation of Equation \eqref{eq:multigauss} for the total spectrum~$T_b$ and the $i$th component $T_{b, i}$.
When the component is strongly blended, $f_{\rm Distinct}$ is small since the shape is not apparent in the total spectrum and its 2nd derivative will not be strongly peaked.
On the other hand, a distinct peak in the total spectrum will have a large $f_{\rm Distinct}$ as both derivatives will contain a similar local minimum.
In the trivial case of a single Gaussian component, $f_{\rm Distinct}=1$.
We classify each component as distinct or blended using this criterion.

\subsection{Number of components recovered}
\label{appsub:mgauss_numberofcomponents}

Our key results in this paper rely on the $\Delta {\rm BIC}$ comparison, where the ${\rm BIC}$ value for each fit depends on the number of free parameters.
Since the multi-Gaussian model is flexible, as the number of components can vary, it is important for the fitting method to broadly recover the correct number of components.
These issues are discussed in depth in several works \citep[e.g.,][]{Lindner2015AJ....149..138L,RienerKainulainen2020A&A...633A..14R}.

\begin{figure*}
\includegraphics[width=\textwidth]{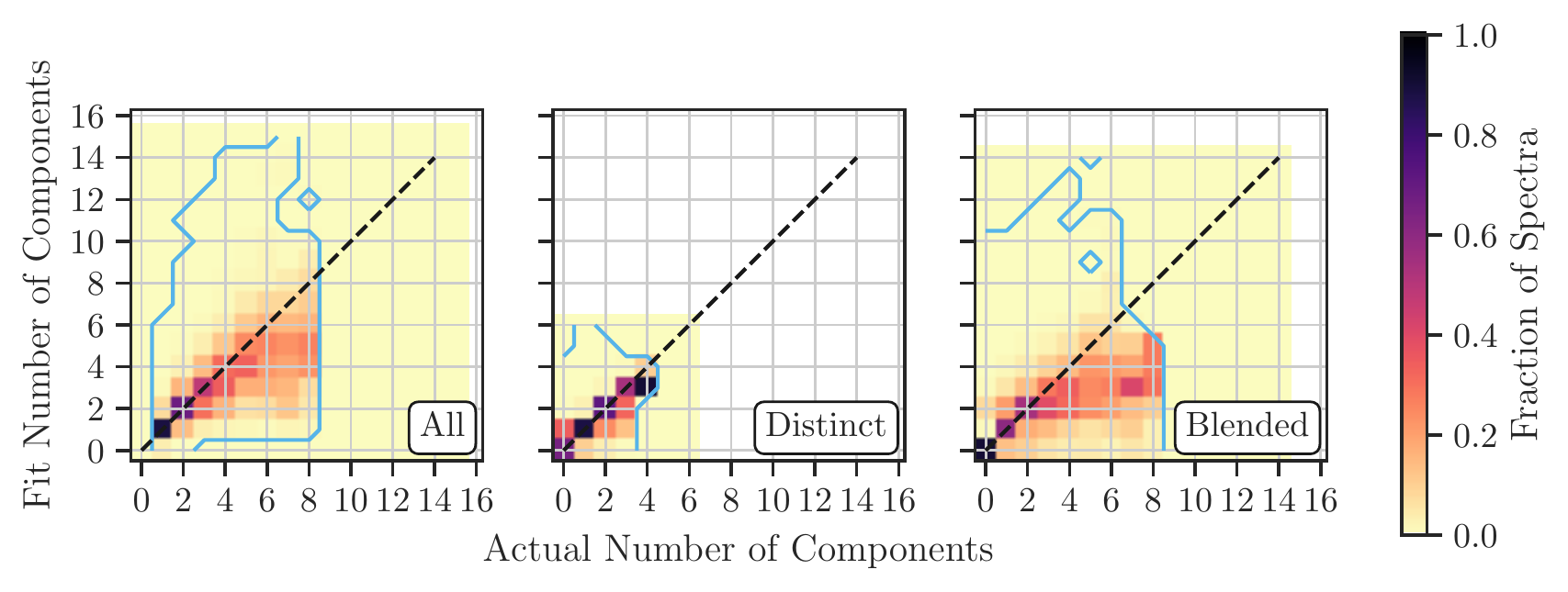}
\vspace*{-10px}
\caption{
\label{fig:ncomp_recovery}
Confusion matrices comparing the actual versus fitted number of Gaussian components from the synthetic spectra for all (left), distinct (centre) and blended (right) components.
The values shown are fractions normalized by column, and the cyan contours indicate the location of all non-zero values.
We find a tendency to underestimate the true number of all components for complex spectra ($N_{\rm comp} > 4$), however, this is primarily due to blended components which are poorly distinguished in the total model.
In contrast, the number of distinct components is well-recovered in the fits.
We conclude that our multi-Gaussian fitting is not systematically biased, thereby strengthening the comparison with the opaque model (Equation~\ref{eq:thickhi_profile}).
}
\end{figure*}

Figure~\ref{fig:ncomp_recovery} shows the actual versus recovered number of Gaussian components for all, distinct, and blended components.
We find a broad range in the number of components identified by the fitting routine, though for all components, there is a bias to underestimate the true number of components for complex spectra ($N_{\rm comp} > 4$).
By splitting the components into distinct and blended sets, we find that this bias for complex spectra is due to blended components, which are difficult to fully recover from a single spectrum with no additional information.
On the other hand, our fitting method performs well for recovering distinct components.

These results demonstrate that the number of components identified by our multi-component Gaussian fitting method is not systematically biased, consistent with previous works using this method \citep{Lindner2015AJ....149..138L,MurrayStanimirovic2017ApJ...837...55M,RienerKainulainen2020A&A...633A..14R}, and supports our conclusions for the $\Delta {\rm BIC}$ model selection in \S\ref{sub:model_tests}.

\subsection{Recovery of Gaussian components}
\label{appsub:mgauss_recovery}

The investigation of recovering Gaussian component parameters, both in these tests and applied to the observations, is beyond the scope of this work and will be presented in E.~W.\ Koch et al.\ (in preparation).
Here, we note two conclusions regarding the recovery of the correct component parameters:
\begin{enumerate}
    \item By comparing to the uniform distribution of drawn parameters (Equation~\eqref{eq:app_synth_distribs}), we find broad agreement over most of the chosen parameter space.
    The recovery fraction decreases for (i) extremely wide ($>40$~\kms; usually highly blended) components and (ii) for narrow ($<1.2$~\kms) components with ${\rm S/N}<5$.
    
    \item Narrower components have higher recovery fractions as they are more likely to be classified as distinct (Equation~\eqref{eq:distinct_def}) and will less overlap with other components.
    This introduces a bias that may affect the average component line width and requires further exploration.
    
\end{enumerate}

While the above points affect the component distributions of the multi-Gaussian fits, they do not strongly influence the overall fit and so do not affect our model selection results.

\label{lastpage}
\end{document}